\documentclass[useAMS,usenatbib]{mn2e}

\usepackage{epsfig}
\usepackage{myaasmacros}
\usepackage{amsmath}
\usepackage{bm}
\usepackage{amsfonts}
\usepackage{subfig}
\usepackage{comment}
\usepackage{color}
\usepackage{setspace}
\usepackage{fancyhdr}
\usepackage[shortlabels]{enumitem}

\title[Stellar feedback in the {\sc Gaea} model] 
{Galaxy assembly, stellar feedback and metal enrichment: the view from
  the {\sc Gaea} model}

\author[Hirschmann et al.]{Michaela Hirschmann$^{1,2}$\thanks{E-mail:
hirschma@iap.fr}, Gabriella De Lucia$^{2}$, Fabio
Fontanot$^{2}$\\
$^{1}$Sorbonne Universit\'es, UPMC-CNRS, UMR7095, Institut d'~Astrophysique de Paris, F-75014
Paris, France\\
$^{2}$INAF - Astronomical Observatory of Trieste, via G.B. Tiepolo 11,
I-34143 Trieste, Italy}

\begin{document}

\date{Accepted ???. Received ??? in original form ???}

\pagerange{\pageref{firstpage}--\pageref{lastpage}} \pubyear{2002}

\maketitle

\label{firstpage}

\begin{abstract}
One major problem of current theoretical models of galaxy formation is
given  by their inability to reproduce the apparently
`anti-hierarchical' evolution of   galaxy assembly: massive galaxies
appear to be in place since $z\sim 3$,  while a significant increase
of the number densities of low mass galaxies is  measured with
decreasing redshift. In this work, we perform a systematic  analysis
of the influence of different stellar feedback schemes, carried out
in the framework of \textsc{Gaea}, a new semi-analytic model of galaxy
formation. It includes a self-consistent treatment for the timings of
gas, metal and energy recycling, and for the chemical yields. We show
this to be crucial to use observational  measurements of the
metallicity as independent and powerful constraints for the adopted
feedback schemes. The observed trends can be reproduced  in the
framework of either a strong ejective or preventive feedback model. In
the former case, the gas ejection rate must decrease significantly
with cosmic time (as suggested by parametrizations of the cosmological
``FIRE'' simulations). Irrespective of the feedback scheme used, our
successful models always imply that up to 60-70 per cent of the
baryons reside in an `ejected' reservoir and are unavailable for
cooling at high redshift. The same schemes predict physical properties
of model galaxies (e.g. gas content, colour, age, and metallicity)
that are in much better agreement with observational data than our
fiducial model. The overall fraction of passive galaxies is found to
be primarily determined  by internal physical processes, with
environment playing a secondary role.  
\end{abstract}

\begin{keywords}
galaxies: evolution - galaxies: formation
\end{keywords}

%******************************************************************************
%******************************************************************************
\section{Introduction}\label{intro}
%******************************************************************************
%******************************************************************************

Within the framework of the currently favoured dark energy dominated
dark  matter paradigm ($\Lambda$CDM), large-scale structure develops
from small-scale density fluctuations, in a bottom-up fashion driven
by gravitational forces (\citealp{Blumenthal84}). The baryonic
component follows the evolution of dark matter: gas that remains
trapped in the potential wells of dark matter haloes is shock heated
to the virial temperature, and later cools and condenses in a disk,
where it forms stars. These eventually explode as supernovae (SN),
heating the surrounding gas and driving spectacular galactic outflows 
(\citealp{White78, Rees77}). Our understanding of the various baryonic  
processes at play is far from complete, and complicated by the fact
that they are entangled in a complex network of actions,
back-reactions and self-regulations. 

One naive expectation of the hierarchical scenario is that the
assembly history of galaxies should parallel that of their parent dark
matter haloes: most massive galaxies should assemble {\it later} than
their lower mass counterparts, and this should manifest in a
significant evolution of the galaxy stellar mass function (GSMF). 
  In fact, a number of observational studies  have highlighted an
  `anti-hierarchical' (because in marked contrast with these
  expectations) trend in the evolution of this basic metric
  (e.g. \citealp{Fontana04, Pozzetti07, Marchesini09, Muzzin13} and  
  references therein): the low mass end of the observed GSMF appears 
  to evolve more rapidly than the high mass end, i.e. the most massive 
  galaxies assemble {\it earlier} than their lower mass counterparts.
While the discrepancy at the massive end is heavily affected by
uncertainties in the measurements of galaxy stellar masses, that at
low masses is robust and has proved to be of difficult solution in the
framework of hierarchical model of galaxy formation
\citep[e.g.][]{Fontanot09, Weinmann12}. When including strong stellar
feedback, hierarchical models are able to reproduce the observed GSMF
in the local Universe, but systematically over-predict the number
densities of sub-M$_*$ galaxies at higher redshift ( e.g.
  \citealp{Guo11}, see also Fig. 4 in \citealp{Somerville15}). What
appears to be needed is a mechanism that is able to decouple the
growth of low-mass galaxies (that occurs late) from that of their host
dark matter haloes (that occurs early). The obvious suspect is {\it
  feedback}, from massive stars and/or from Active Galactic Nuclei
(AGN). 

Observational signatures in broad emission/absorption lines of AGN
(e.g. \citealp{Moe09,  Dunn10, Maiolino12, Cicone14, Brusa15,
  Harrison14}) and cavities in the hot, X-ray emitting gas surrounding
radio galaxies (e.g.  \citealp{McNamara07, McNamara11, Cattaneo09,
  Fabian12}) highlight the presence of radiative and mechanical
feedback from central accreting black holes. These powerful sources of
energy have long been suspected to play an important role in the
`cooling-flow' problem. Early models accounted for this form of
feedback by simply assuming that cooling would be suppressed above
some critical halo mass or circular velocity
\citep[e.g.][]{Kauffmann99, DeLucia04}. More 
sophisticated schemes have been implemented in the last decade, both
in semi-analytic models of galaxy formation and hydrodynamical
simulations (e.g. \citealp{Springel05a, Croton06, Bower06, Choi15,
  Steinborn15} and references therein). While these are  successful
(under certain conditions) in offsetting the cooling flow in massive
haloes and lead to massive central galaxies dominated by 
old stellar populations, some problems remain. E.g. the predicted AGN
activity levels of local galaxies exhibit trends as a function of
stellar mass and halo mass that are in contrast with those measured
(\citealp{Fontanot11}), and the metallicities predicted for
the most massive galaxies are significantly lower than those observed
\citep{DeLuciaBorgani12}. Some cosmological simulations also tend to
over-estimate the massive end of the GSMF due to an inefficient
suppression of star formation at the centre of galaxy groups and 
clusters (\citealp{Puchwein13, Khandai15, Hirschmann14b, Genel14}). 

Observations also reveal large, supernova-driven, bi-polar galactic
outflows with velocities up to 500~km/s in nearby galaxies
(e.g. \citealp{Heckman90, Heckman00, Martin05,
  Veilleux05, Rubin14}), and in high-z star-forming objects
(e.g. \citealp{Steidel96, Pettini00, Weiner09, Martin13,
  Newman13}). These stellar-driven winds are expected to  play an
important role in regulating star formation in galaxies at and 
below the exponential cut-off of the GSMF. The energy release
originates from a combination of exploding SN type II and Ia, and from
massive stars in form of ionising radiation, radiation pressure and
stellar winds. The details of the  energetics and outflows, as well as
the fate of the outflowing gas, are not well understood neither from
the observational, nor from the theoretical point of view. As a
consequence, modelling of this particular physical process is
necessarily schematic. Nevertheless, theoretical studies show
that galactic winds affect significantly a number of galaxy physical
properties such as their star formation rates, metal and gaseous
content, size, gas kinematics as well as the stellar mass assembly
(ratio of in-situ formed to accreted stars), and the baryonic and metal
content of the inter-galactic and intra-cluster medium
(e.g. \citealp{DeLucia04}, \citealp{ Dave11a,  Hirschmann12a,
  Lackner12, Aumer13, Hopkins14, Hirschmann15}).  

It is interesting to note that the above mentioned over-production of
sub-M$_*$ galaxies is connected to a long-standing problem in
hydrodynamical simulations: gas cools efficiently at high redshift and
in small compact haloes. The originating clumps merge via dynamical
friction leading to a significant  transfer of angular momentum from
baryons to the dark matter. This results in spiral galaxies with
relatively large bulges and compact disks (the `angular momentum
catastrophe'). Significant effort has been devoted to improve the
adopted stellar feedback scheme in hydrodynamical simulations
\citep[e.g.][]{Stinson13, Aumer13,   Schaye14, Hopkins14,
  Murante15}. These studies are now able to obtain baryon conversion
efficiencies close to those expected for $\sim$M$_*$ galaxies. Very
recent cosmological hydrodynamical simulations have also been
successfully `calibrated' to reproduce the observed GSMF in the local
Universe (e.g. the Illustris and the EAGLE simulations -
\citealp{Genel14, Furlong15}. The latter also successfully reproduce
the observed evolution of the GSMF). 

Significant efforts to improve the adopted stellar feedback scheme
have also been made in the framework of semi-analytic models of galaxy
formation (\citealp{Knebe15}), and a few schemes have been shown to
successfully reproduce the observed evolution of the GSMF
\citep{Lagos13, Henriques13, White15, Lu15}. Albeit representing
important improvements with respect to previous results, none of these
models appears to be completely satisfactory with respect to e.g. the
predicted colour distributions (\citealp{Henriques13}) or metal evolution
(\citealp{White15}). All these models are based on the instantaneous
recycling approximation, where energy and metals are assumed to be
 released   immediately after star formation, i.e. neglecting the
finite life-time of stars and its dependence on stellar mass. This
represents an important limitation, and prevents the use of
observational data on the gaseous and stellar metallicity content of
galaxies as independent constraints. 

In this work, we will carry out a systematic analysis of the influence of
different stellar feedback models, with a focus on the
mechanisms/schemes required to reproduce the observed evolution of the
GSMF. Our work is based on an evolution of the semi-analytic model
presented in \citet{DeLucia07}, and includes the detailed chemical
enrichment scheme introduced in \citet{DeLucia14}. Although close to
the latest developments of the {\sc  L-Galaxies} model \citep{Guo11,
  Henriques15} in terms of basic structure and modelling for some
specific physical processes, our model has differentiated
significantly from this parallel branch for the treatment of various
processes and modification of the chemical enrichment and feedback
scheme. To emphasise these differences, we will refer in the following
to our updated model as the {\sc GAEA}\footnote{Named after Gaea, one
  of the Greek primordial deities. Personification of the Earth and
  creator of the Universe.} (GAlaxy Evolution and Assembly)
model\footnote{Note that galaxy catalogues for the {\sc GAEA} model
  will be made publicly available under
  http://gavo.mpa-garching.mpg.de/Millennium/}. 

Our paper is structured as follows. In section \ref{model} we describe
the main features of our {\sc Gaea} model, and describe in detail the
different feedback schemes that we have tested. Sections
\ref{downsizing} and \ref{origin} focus on the impact of stellar
feedback on the predicted evolution of the GSMF, and analyse the
corresponding predictions as for the ejection and recycling rates of
gas. We then discuss the influence of the same feedback schemes on the
gaseous and metal content of galaxies, colour and star formation rate
distributions, and stellar populations in sections \ref{Gas},
\ref{SFR} and \ref{Stellarpop}. Finally, we discuss our results in
Section \ref{discussion}, and summarise our findings in Section
\ref{summary}.

%******************************************************************************
%******************************************************************************
\section{The {\sc Gaea} galaxy formation model }\label{model}
%******************************************************************************
%******************************************************************************

Our galaxy formation model corresponds to that described in
\citet{DeLucia14}, with modifications of the stellar feedback scheme
and of some of the model parameters, as described in the following
sections. The model is based on that described in detail by
\citet[][and references therein]{DeLucia07}, but has been modified to
follow more accurately processes on the scale of the Milky Way
satellites as described in \citet{DeLucia08} and \citet{Li10}. In
particular, our model assumes an earlier re-ionisation: it is
  assumed to start at 
  $z=15$ and to last for about 0.12 Gyr, while in the De Lucia \&
  Blaizot model it was assumed to start at $z=8$ and to last
  approximately for the same amount of time. In addition, we assume
  that gas cooling is suppressed below $10^4$~K, while in the original
  model, haloes with virial temperature lower than $10^4$~K were able
  to cool as much gas as a $10^4$~K halo with the same
  metallicity. Finally, we assume that for galaxies residing in haloes
  with virial mass below $5 \times 10^{10} M_\odot$, most of the new
  metals (95 per cent) are ejected directly into the hot gas phase,
  while in the De Lucia \& Blaizot model all new metals were
  immediately mixed with the cold gas in the disk. The models
    discussed in this paper, also differ from that adopted in
    \citet{DeLucia07} for the modelling assumed to estimate disk radii
    and the critical surface  density below which no star formation
    takes place. For details, we  refer to   \citet{DeLucia08}. We 
  stress   that, while these modifications are important at the scale
  of Milky Way   satellites \citep{Li10}, they have little or no
  influence on the results shown in this study.

In the {\sc Gaea} model, the evolution of baryons is traced in four
different reservoirs, the stellar component of galaxies, the cold gas
in galaxy discs, the hot gas associated with the dark matter haloes,
and the ejected gas component. To model the mass and energy transfer
between these different reservoirs, the {\sc Gaea} model includes
physically and/or observationally motivated prescriptions for gas
cooling, re-ionisation, star formation, stellar feedback and gas
recycling, metal evolution, black hole growth, AGN feedback, disk
instabilities and environmental effects.  For most of these
prescriptions, we follow the same approach as in \citet{DeLucia07}
 except for the   modifications discussed above, and with the
major exception of chemical enrichment (as described in
\citealp{DeLucia14}), and stellar feedback and gas recycling (which
are the subject of this study). Below, we describe the simulations
used in this study, give a summary of our chemical enrichment model,
and describe in detail the stellar feedback schemes tested in this
work. 

%******************************************************************************
\subsection{Dark matter simulation and cosmology}\label{Sim}
%******************************************************************************

In this study, we take advantage of the dark matter merger trees
extracted from the Millennium Simulation (\citealp{Springel05n}). The
simulation assumes a WMAP1 cosmology with $\Omega_{\Lambda}$ = 0.75,
$\Omega_{\mathrm{m}}$ = 0.25, $\Omega_{\mathrm{b}}$ = 0.045, n=1,
$\sigma_8$ = 0.9, and h=0.73. To test the convergence of our results
with resolution, we also run the {\sc Gaea} model over the merger
trees of the Millennium-II Simulation \citep{Boylan-Kolchin09}. 
This simulation adopts the same cosmology of the Millennium, but
corresponds to a smaller box ($100\,{\rm Mpc}\,{\rm h}^{-1}$ against
$500\,{\rm Mpc}\,{\rm   h}^{-1}$), five times better spatial
resolution (the Plummer equivalent softening of the Millennium II is
$1.0\,{\rm kpc}\,{\rm h}^{-1}$) and 125 times better mass resolution
(the particle mass is $6.9 \times 10^6\,{\rm   M}_{\odot}{\rm
  h}^{-1}$). Convergence tests are presented in Appendix
\ref{resolution}.  

Note that more recent measurements of the cosmological parameters
suggest a  lower value for $\sigma_8$ (the latest results from the
Planck Collaboration suggest a value of $0.831\pm0.013$). This means 
that the local Universe from the Millennium and Millennium II
simulations could be more evolved than the real one \citep[see
e.g.][]{Wang08}. We have explicitly tested to what
extent our results change when adopting a lower value for
$\sigma8$. In particular, we used the simulation described in
\citet{Wang08} that adopts $\sigma8=0.7$. Confirming results from that
paper (see also \citealt{Guo13, Gonzalez14}), we find that model
parameters need  to be re-tuned to reproduce the same level of
agreement at $z=0$, but results are qualitatively the same. 

%******************************************************************************
\subsection{The chemical evolution scheme}\label{Chemistry}
%******************************************************************************

The interstellar medium (ISM) of galaxies is enriched by metals
released by stellar winds and by SN explosions. The large majority of
published work is based on SAMs that assume an instantaneous recycling
approximation, i.e. they neglect the dependence of stellar lifetimes
on stellar mass, and assume that gas recycling and chemical enrichment
occur `instantaneously' (i.e. within the individual time-step of the
code). Our {\sc Gaea} model adopts the chemical enrichment scheme
described in \citet{DeLucia14}, explicitly accounting for the finite
life times of stars, and tracing individual element abundances. While
there are some uncertainties in the adopted chemical yields and SN Ia
model, relaxing the instantaneous recycling approximation prevents us
from considering the metal yield as a free adjustable parameter of the
model.  As we will discuss in the following, this allows us to use
observational data on chemical abundances as independent constraints,
and therefore draw more robust conclusions on the gas and metal
recycling scheme.  

For details on the implementation, we refer to \citet{DeLucia14}. The
model assumes a Chabrier IMF \citep{Chabrier03} and stellar life times
by \citet{Padovani93}. Stars with masses below $8\,{\rm M}_\odot$ are
assumed to enrich the ISM mainly through AGB winds, while more massive
ones are expected to die as SNII. In our chemical enrichment scheme,
the only free parameter is represented by the realisation probability
of a given SNIa scenario, and is constrained by the observed Fe
content of the Milky Way disk (we refer to \citealp{DeLucia14} for
details). Our fiducial model assumes a delay time distribution (DTD)
corresponding to the single degenerate scenario described in
\citet{Matteucci01} and \citet{Bonaparte13}, and the metal yields of
\citet{Thielemann03} for SNIa, of \citet{Chieffi02} for SNII, and
those of \citet{Karakas10} for winds from low- and intermediate mass
stars. 

\citet{DeLucia14} have applied this chemical scheme to a set of
simulated Milky-Way size haloes (those completed within the Aquarius
project - \citealp{Springel08}). They also discuss the dependence on
various model ingredients, and the differences with respect to the
same physical model assuming an instantaneous recycling
approximation. Model predictions were shown to be in fair agreement
with the observed chemical composition of the Milky Way, in particular
of its disk component, and of its satellites. 

When applying this model to a cosmological volume, we found that we
had to adjust some model parameters in order to match the
normalisation of the mass-metallicity relation as measured from the
Sloan Digital Sky Survey (SDSS). We discuss how these adjustments
affect predictions for Milky Way like galaxies in appendix
\ref{MWappendix}. In a recent paper, \citet{Yates13} have
independently included an updated chemical enrichment model into the
semi-analytic model by \citet{Guo11}. They show that their
implementation is able to reproduce simultaneously the chemical
content of Milky Way-like galaxies and the observed mass-metallicity
relation in the local Universe. They do not discuss, however, how
their updated chemical model affects some basic predictions like the
galaxy luminosity or stellar mass function and their   evolution (some
differences are expected as they also modify the adopted stellar
feedback). As we will show in the following, we have identified a
stellar feedback scheme that is able to match the local galaxy mass
function and its evolution to higher redshift, as well as the
metallicity content of galaxies, down to the scale of the ultra-faint
dwarfs of the Milky Way. 

%\begin{spacing}{\baselinestretch}
\begin{table*}
\centering
\begin{tabular}{ | p{1.5cm} || p{5.cm} p{5.cm} p{3.5cm} p{0.001cm} |} 
\centering{Model (color)} & \centering{Re-heating} [$\dot{M}_{\mathrm{reheat}}$]&
\centering{Ejection} [$\dot{M}_{\mathrm{eject}}$] &
\centering{Re-incorporation} [$\dot{M}_{\mathrm{reinc}}$] & \\ \hline \hline

\centering{1. Fiducial (lila)}  & \centering{$\frac{4}{3}
\epsilon_{\mathrm{reheat}} \cdot \frac{E_{\mathrm{tot,SN}}}{V_{\mathrm{vir}}^{2}} \cdot
\dot{M}_{\mathrm{star}}$, $\bm{\epsilon_{\mathrm{reheat}} = 0.02}$} &
\centering{Centrals: all re-heated gas ejected,  Satellites: re-heated
  gas added to hot halo gas} & \centering{$ \gamma
\frac{{M}_{\mathrm{eject}}}{t_{\mathrm{dyn}}}$, $\bm{\gamma = 0.1}$} & \\ 
\hline 

\centering{2. Lag13 (dark blue)} & \centering{$ \epsilon_{\mathrm{reheat}}
  \left(\frac{\Sigma_{\mathrm{gas}}}{1600\,{\rm M}_\odot
      \mathrm{pc}^{-2}}\right)^{-0.6} \cdot 
\left(\frac{f_{\mathrm{cold}}}{0.12}\right)^{0.8}  
\dot{M}_{\mathrm{star}}$, $\bm{\epsilon_{\mathrm{reheat}} = 0.2}$} &
\centering{All re-heated gas ejected} &  \centering{$ \gamma
\frac{{M}_{\mathrm{eject}}}{t_{\mathrm{dyn}}}$, $\bm{\gamma = 0.1}$} &  \\ 
\hline 

\centering 3. Hop12 (light blue) &\centering $\epsilon_{\mathrm{reheat}} \cdot 
\left( \frac{\Sigma_{\mathrm{gas}}(r)}{10\,{\rm M}_\odot 
    \mathrm{pc}^{-2}}\right)^{-0.5}  \left(
  \frac{V_{\mathrm{circ}}(r)}{100 \rm{km}\,{\rm s}^{-1}}\right)^{-1.1}  
\cdot \dot{M}_{\mathrm{star}}$, $\bm{\epsilon_{\mathrm{reheat}} = 1.0}$ &
\centering{Centrals: all re-heated gas ejected,  Satellites: re-heated
  gas added to hot halo gas} & \centering $ \gamma
\frac{{M}_{\mathrm{eject}}}{t_{\mathrm{dyn}}}$,  $\bm{\gamma = 0.1}$ &
\\ \hline 

\centering 4. Guo11 (turquoise) & \centering $\epsilon_{\mathrm{reheat}} \cdot
\left[ 0.5 + \left( 
    \frac{V_{\mathrm{max}}}{70\ \mathrm{km/s}} \right)^{-3.5} \right]
\cdot \dot{M}_{\mathrm{star}}$, $\bm{\epsilon_{\mathrm{reheat}} = 0.7}$ &
\centering $ \frac{\dot{E}_{\mathrm{FB}} - 0.5   \dot{M}_{\mathrm{reheat}}
  V_{\mathrm{vir}}^2}{0.5 V_{\mathrm{vir}}^2}$ with
$\dot{E}_{\mathrm{FB}} = \epsilon_{\mathrm{eject}} \cdot \left[ 0.5 +
  \left(  \frac{V_{\mathrm{max}}}{70\ \mathrm{km/s}} 
  \right)^{-3.5}\right] \cdot 0.5  \dot{M}_{\mathrm{star}}
V_{\mathrm{SN}}^2$, 
$\bm{\epsilon_{\mathrm{eject}} = 0.35}$ & \centering $\gamma
\frac{{M}_{\mathrm{eject}}}{t_{\mathrm{dyn}}}$,  $\bm {\gamma = 0.5}$ &\\
\hline  

\centering 5. Hen13 (green) & \centering $\epsilon_{\mathrm{reheat}} \cdot
\left[ 0.5 + \left( 
    \frac{V_{\mathrm{max}}}{336\ \mathrm{km/s}} \right)^{-0.46}
\right] \cdot \dot{M}_{\mathrm{star}}$, $\bm{\epsilon_{\mathrm{reheat}} = 1.0}
$ & \centering As in Guo11 with $\dot{E}_{\mathrm{FB}} =
\epsilon_{\mathrm{eject}} \cdot \left[ 0.5 +  \left(
    \frac{V_{\mathrm{max}}}{405\ \mathrm{km/s}} \right)^{-0.92}\right]
\cdot 0.5  \dot{M}_{\mathrm{star}} V_{\mathrm{SN}}^2$,
$\bm{\epsilon_{\mathrm{eject}} = 0.25 }$ &
\centering $ \gamma
\frac{{M}_{\mathrm{eject}}}{t_{\mathrm{reinc}}},\ \mathrm{with}\
t_{\mathrm{reinc}} = \frac{10^{10}M_\odot}{M_{\mathrm{vir}}}$,  $\bm{\gamma
= 1.0}$ &\\ 
\hline

\centering 6. zDEP (yellow) & \centering
$\epsilon_{\mathrm{reheat}} \cdot \left[ 0.5 + 
  (1+z)^3 \left( \frac{V_{\mathrm{max}}}{70\ \mathrm{km/s}}
  \right)^{-3.5} \right] \cdot \dot{M}_{\mathrm{star}}$,
$\bm{\epsilon_{\mathrm{reheat}} = 0.7}$ & \centering As in Guo11 with
$\dot{E}_{\mathrm{FB}} = \epsilon_{\mathrm{eject}}  \cdot \left[ 0.5 +
  (1+z)^3 \left( \frac{V_{\mathrm{max}}}{70\ \mathrm{km/s}}
  \right)^{-3.5}\right] 
\cdot 0.5 \dot{M}_{\mathrm{star}} V_{\mathrm{SN}}^2$,
 $\bm{\epsilon_{\mathrm{eject}} = 0.15}$ &  \centering $
\gamma \frac{{M}_{\mathrm{eject}}}{t_{\mathrm{dyn}}}$, $\bm{\gamma = 0.5}$
& \\ \hline 

\centering 7. FIRE (orange) & \centering $\epsilon_{\mathrm{reheat}}
(1+z)^{1.25} \cdot \left( 
  \frac{V_{\mathrm{max}}}{60\ \mathrm{km/s}} \right)^{\alpha}  
\cdot \dot{M}_{\mathrm{star}}$, 
$V_{\mathrm{max}} < 60\ \mathrm{km/s}\ \rightarrow\ \alpha = -3.2$,  
$V_{\mathrm{max}} > 60\ \mathrm{km/s}\ \rightarrow\ \alpha = -1.0$,
$\bm{\epsilon_{\mathrm{reheat}} = 0.3}$
  & \centering As in Guo11 with $\dot{E}_{\mathrm{FB}} =
  \epsilon_{\mathrm{eject}} (1+z)^{1.25} \cdot \left(
    \frac{V_{\mathrm{max}}}{60\ \mathrm{km/s}} \right)^{\alpha} \cdot
  0.5 \dot{M}_{\mathrm{star}} V_{\mathrm{SN}}^2$,
  $\bm{\epsilon_{\mathrm{eject}} = 0.1}$ &  \centering  $\gamma
\frac{{M}_{\mathrm{eject}}}{t_{\mathrm{reinc}}},\ \mathrm{with}\
t_{\mathrm{reinc}} = \frac{10^{10}M_\odot}{M_{\mathrm{vir}}}$, $\bm{\gamma
= 1.0}$ & \\ \hline

\centering  8. PREH (red) & \centering{$\frac{4}{3} \epsilon_{\mathrm{reheat}}
  \cdot
  \frac{E_{\mathrm{tot,SN}}}{V_{\mathrm{vir}}^{2}} \cdot
  \dot{M}_{\mathrm{star}}$, $\bm{\epsilon_{\mathrm{reheat}} = 0.02}$} &
\centering $ \dot{M}_{\mathrm{gas, preheat}} +
\dot{M}_{\mathrm{reheat}} $ with  $ \dot{M}_{\mathrm{gas, preheat}} =
(1-f_{\mathrm{preheat}}) f_{\mathrm{bar}} 
\dot{M}_{\mathrm{DM, infall}}, f_{\mathrm{preheat}} =
\frac{M_{\mathrm{vir}}}{M_{\mathrm{crit}}}$, $M_{\mathrm{crit}} =
10^{12} M_\odot$;
& \centering
$\gamma
\frac{{M}_{\mathrm{eject}}}{t_{\mathrm{reinc}}},\ \mathrm{with}\
t_{\mathrm{reinc}} = \frac{10^{10}M_\odot}{M_{\mathrm{vir}}}$,  $\bm{\gamma
= 0.5}$ & \\ \hline
\end{tabular}
\caption{Overview of the stellar feedback models and of the
  corresponding free parameters for gas reheating, ejection and
  re-incorporation, tested in the framework of the {\sc Gaea}
  model. In the PREH model, gas that is pre-heated is added to the
  `ejected' reservoir. We refer to section \ref{Stellarfb} for details.}
\label{feedbacktab}
\end{table*}
 %\end{spacing}

%******************************************************************************
\subsection{Stellar feedback models}\label{Stellarfb}
%******************************************************************************

The energy supplied by massive stars in the form of SN and stellar
winds represents the engine that drives the galactic scale outflows
observed in actively star forming galaxies, both in the local Universe
and at high redshift (see Section \ref{intro}). Unfortunately, the
observational measurements available refer to material that is still
relatively deep within the gravitational potential well of the halo so
that it is difficult to translate the estimated outflow rates into
rates at which mass, metals, and energy escape from the galaxies and
are transported into the inter-galactic medium. As a consequence,
galaxy formation models rely on analytic parametrizations or
`sub-grid' models based on our best understanding of the theory and/or
on observational measurements.

In order to test the influence of different scalings, we have
implemented eight different schemes for stellar feedback and gas
recycling in our {\sc Gaea} model. Four of these have been introduced
in the framework of previously published semi-analytic models. 
 For each scheme, we have varied the stellar feedback (re-heating
and ejection) and reincorporation efficiencies (see Table
\ref{feedbacktab}) so as to match the exponential cut-off of the
stellar mass function at $z=0$, and by simultaneously trying to obtain 
a good match with the observed mass-metallicity relation in the local
Universe and the measured evolution of   the galaxy stellar mass
function at higher redshift. All other model parameters have been left
unchanged (so they correspond to those used in \citealt{DeLucia14} and
references therein). As mentioned above, we have modified some model
parameters for our fiducial model (see Appendix A) with respect to
\citealt{DeLucia14} to recover the correct normalization of the
mass-metallicity relation. We want to emphasize that for our fiducial
model, we were not able to find any combination of parameters to match
the observed GSMF and its evolution, highlighting the difficulty in
reproducing the anti-hierarchical trend in galaxy stellar mass growth,
which is the main motivation of this work to explore different models
for stellar feedback.

 Table \ref{feedbacktab} summarises the different stellar feedback  
  schemes discussed and analysed in this work, and lists the
  corresponding parameter values for the stellar feedback and
  reincorporation efficiencies. As detailed below, we assume that
stellar winds and SN are able to reheat some fraction of the cold gas
in the disk. This `reheated' gas is assumed to leave the galaxy, but
remains bound to its parent halo and becomes part of its hot gas
component, unless the energy is large enough to escape the potential
well of the halo. This `ejected' component is stored in a reservoir,
and is not available for cooling until it is `re-incorporated' into
the hot gas component. A feedback scheme is therefore fully
characterised by a reheating rate, an ejection rate and a
re-incorporation rate. Each rate is characterised by some efficiency
($\epsilon_{\mathrm{reheat}}$ and $\epsilon_{\mathrm{eject}}$ for the
reheating and ejection rates) or factor ($\gamma$ for the
re-incorporation). In the following, we describe in detail each of the
schemes tested in this work. In all cases, we assume that when
galaxies are accreted on larger systems (i.e. when they become
satellites), they are instantaneously stripped of their hot and
ejected gas reservoirs. The latter is added to the ejected component
associated with the central galaxy. As a consequence, gas
re-incorporation is only modelled onto central galaxies of haloes. 
 
\subsubsection{Energy-driven winds due to SNae (Fiducial)}\label{fiducial} 
%******************************************************************************

The `fiducial' feedback scheme implemented in \citet{DeLucia14}
corresponds to the `energy-driven' scheme described in
\citet{DeLucia04}. The reheated gas mass rate
$\dot{M}_{\mathrm{reheat}}$ is calculated as: 
\begin{equation}
\dot{M}_{\mathrm{reheat}} = \frac{4}{3} \epsilon_{\mathrm{reheat}}
\frac{E_{\mathrm{tot,SN}}}{V_{\mathrm{vir}}^{2}} \times
\dot{M}_{\mathrm{star}}, 
\end{equation}
where $\dot{M}_{\mathrm{star}}$ is the star formation rate (SFR)  in one
  individual time-step of integration, $V_{\mathrm{vir}}$ the virial velocity
of the halo (subhalo for satellite galaxies), $\epsilon_{\mathrm{reheat}}$ the
re-heating efficiency, and $E_{\mathrm{tot,SN}}$ the total amount of energy
released by SNIa and SNII. The latter quantity is self-consistently calculated
within our non instantaneous recycling scheme, i.e. accounting for finite
stellar lifetimes. We assume that all reheated gas of central galaxies is
ejected out of the halo ($\dot{M}_{\mathrm{eject}}=\dot{M}_{\mathrm{reheat}}$),
while the reheated gas of satellite galaxies is added to the hot gas
content of the parent halo (that can only cool onto the central
galaxy). We have   verified that varying the exponent in the
denominator of Eq.~1 with   values  from $-1$ (e.g. assuming `momentum
driven winds') to $-4$   hardly changes the   results discussed in the
paper for this   particular model. 

A certain fraction of the ejected gas mass is assumed to be
re-accreted back onto the (parent) halo on a halo dynamical time-scale
$t_{\mathrm{dyn}} = R_{\mathrm{vir}}/V_{\mathrm{vir}}$
($R_{\mathrm{vir}}$ and $V_{\mathrm{vir}}$ are the virial radius and
velocity of the parent halo, respectively): 
\begin{equation}
\dot{M}_{\mathrm{reinc}} = \gamma
\frac{{M}_{\mathrm{eject}}}{t_{\mathrm{dyn}}},
\end{equation}
where $\gamma$ is an adimensional parameter.

As mentioned above, we had to modify the model parameters
with respect to those adopted in \citet{DeLucia14} in order to match  
the observed metallicity content of galaxies in the local Universe.
Specifically, we had to slightly reduce the feedback efficiency (0.02
against 0.05 used in \citealp{DeLucia14}) and the re-incorporation
factor (0.1 against 0.5). In appendix \ref{MWappendix}, we discuss
how these parameter modifications affect results for Milky Way-like
galaxies.  

\subsubsection{Propagation of pressurized bubbles due to SNae (Lag13)} 
%******************************************************************************

\citet{Lagos13} discuss a dynamical model for the propagation and 
evolution of pressurised SNae bubbles in a multiphase ISM. In 
particular, they derive parametrizations for the mass loading of the
cold gas, i.e. the amount of cold gas that is expelled out of the disk
and is thus temporarily not available for cooling. Specifically, they
argue that the often assumed  dependence on the circular velocity or
virial velocity alone is a poor description of the outflow
process. Instead, they propose a dependence on the gas surface density
$\Sigma_{\mathrm{gas}}$ and on the cold gas fraction
$f_{\mathrm{cold}}$ of the galaxy. In particular, their
parametrization of the outflow rate  (reheating rate in our
  terminology)
reads as: 
\begin{equation}
\dot{M}_{\mathrm{reheat}} = \epsilon_{\mathrm{reheat}}
\left(\frac{\Sigma_{\mathrm{gas}}}{1600\,{\rm M}_\odot
    \mathrm{pc}^{-2}}\right)^{-0.6}
\left(\frac{f_{\mathrm{cold}}}{0.12}\right)^{0.8}   
\dot{M}_{\mathrm{star}}, 
\end{equation}
where $\epsilon_{\mathrm{reheat}}$ is the reheating
  efficiency. The 
gas surface density is computed as follows:
\begin{equation}\label{GasSurf}
\Sigma_{\mathrm{Gas}} = \frac{M_{\mathrm{cold}}}{\pi
  R_{\mathrm{gas}}^2}\ \mathrm{ with}\
R_{\mathrm{gas}} =
\frac{J_{\mathrm{gas}}}{2M_{\mathrm{cold}}V_{\mathrm{max}}},
\end{equation}
where $R_{\mathrm{gas}}$ is the scale radius of the gaseous disk,
$M_{\mathrm{cold}}$ the cold gas mass, $J_{\mathrm{gas}}$ the angular
momentum of the gas, and $V_{\mathrm{max}}$ the maximum circular
velocity. We assume the angular momentum of the gas and the
  maximum circular velocity to  be equal to those of the parent dark
  matter (sub)halos (the same assumptions   are made for other models
  that use the same quantities). Following the   implementation of
  this model in the GALFORM semi-analytic code   (\citealp{Lagos13}),
  we assume the reheated gas (both from central and   satellite
  galaxies) to be ejected so that it is not available for cooling until
it is again reincorporated into the hot gas component. We model gas 
re-incorporation as in the fiducial model described above (this
implies that the re-incorporated gas will be able to cool only on the
central galaxy at later times, i.e. gas ejected from satellite
galaxies will never be available to the same galaxies at any later
time). \citet{Lagos13} tested this feedback scheme in the framework of
the GALFORM model, and found that it predicts optical and NIR
luminosity functions in good agreement with data, both at low and high
redshifts, in particular with respect to the shallow measured faint
end slopes. 

\subsubsection{Stellar-driven galactic-scale winds parametrized from
  isolated galaxy simulations (Hop12)}
%******************************************************************************

\citet{Hopkins11} and \citet{Hopkins12a} performed high-resolution
simulations  of isolated galaxies including a new model for stellar
feedback. The key physical processes considered include heating from
both type I and type II supernovae, stellar winds from AGBs, heating
from the shocked stellar winds,  HII photoionisation, and radiation
pressure. In \citet{Hopkins12b}, they parametrize the resulting gas
outflow rates, finding they are well approximated by a momentum-driven
wind scaling ($\sim V_{\mathrm{circ}}^{-1.1}$) with an additional
dependence on the gas surface density $\Sigma_{\mathrm{gas}}$. The gas
re-heating rate $\dot{M}_{\mathrm{reheat}}$ (gas ejection out of the
galaxy), depending on the radial distance from the galaxy centre $r$,
is given by: 
\begin{equation}
\dot{M}_{\mathrm{reheat}} = \epsilon_{\mathrm{reheat}}
\left( \frac{\Sigma_{\mathrm{gas}}(r)}{10\,{\rm M}_\odot
    \mathrm{pc}^{-2}}\right)^{-0.5}  \left(
  \frac{V_{\mathrm{circ}}(r)}{100 \rm{km}\,{\rm s}^{-1}}\right)^{-1.1}  
\dot{M}_{\mathrm{star}},
\end{equation}
where $\epsilon_{\mathrm{reheat}}$ is the re-heating efficiency. 
For simplicity, we use the maximum circular velocity and the gas
surface density measured at the scale radius of the gaseous disk, as
computed for the Lag13 model (see eq.~\ref{GasSurf}).
Like in the fiducial model, we assume that all gas re-heated from
centrals galaxies leaves the halo and is temporarily unavailable for
cooling, while re-heated gas from satellite galaxies is assumed to
remain associated with the hot gas within the parent dark matter
halo. Gas re-incorporation is modelled as in the fiducial model. This
particular model has not yet been tested in the
framework of any semi-analytic model. 

\subsubsection{Strong stellar feedback for dwarf galaxies  (Guo11)} 
%******************************************************************************

The galaxy formation model of \citet{Guo11} is based on that of
\citet{Croton06} and \citet{DeLucia07}, with some important
modifications relative to the modelling of satellite galaxies, and
stellar feedback. In order to reproduce the shallow slope of the
observed present-day stellar mass function, these authors modified the
stellar feedback scheme of \citet{DeLucia07} allowing higher ejection
efficiencies in dwarf galaxies. In particular, the reheated gas mass
rate $\dot{M}_{\mathrm{reheat}}$ is computed as:
\begin{equation}
\dot{M}_{\mathrm{reheat}} = \epsilon_{\mathrm{reheat}} \left[ 0.5 +
\left( \frac{V_{\mathrm{max}}}{70\ \mathrm{km/s}} \right)^{-3.5} \right]
\times \dot{M}_{\mathrm{star}},
\end{equation}
The rate of energy injection by massive stars into disc and halo gas
is parametrized as:
\begin{equation}
\dot{E}_{\mathrm{FB}} = \epsilon_{\mathrm{eject}} \left[ 0.5 +
\left( \frac{V_{\mathrm{max}}}{70\ \mathrm{km/s}} \right)^{-3.5}\right]
\times 0.5  \dot{M}_{\mathrm{star}} V_{\mathrm{SN}}^2, 
\end{equation}
where $0.5 V_{\mathrm{SN}}^2$ is the mean kinetic energy of SN ejecta
per unit  mass of stars formed. In our model, the latter quantity is
self-consistently calculated following our non instantaneous chemical
enrichment scheme. The additional scaling with $(0.5 +
V_{\mathrm{max}}^{-3.5})$ is motivated by the observation that dwarf
galaxies have lower metallicities and less dust than their more
massive counterparts. Therefore, it is plausible that radiative losses
during the thermalisation of ejecta are substantially smaller than in
more massive systems. 

Given this energy input into the disc and halo gas, the total amount
of material that can be escape the halo can be estimated following
energy conservation arguments:
\begin{equation}\label{Encons}
 \dot{M}_{\mathrm{eject}} = \frac{ \dot{E}_{\mathrm{FB}} - 0.5 
 \dot{M}_{\mathrm{reheat}} V_{\mathrm{vir}}^2}{0.5 V_{\mathrm{vir}}^2},
\end{equation}
where $V_{\mathrm{vir}}$ is the escape velocity from the parent halo
(or subhalo for satellite galaxies). 

An important difference of this feedback scheme compared to the
fiducial model, is that all reheated gas from centrals and satellites
is initially added to the hot gas component of the parent halo. Then,
the ejection rate is computed from this updated hot gas reservoir,
that also contains gas heated by e.g. AGN feedback and newly infalling
gas. As consequence, the ejected gas can be larger than the reheated
(by stellar feedback alone) gas.  As we will see below, this will have
important consequences on the predicted star formation rates.  

In the context of our {\sc Gaea} model, we adopt the same gas
re-incorporation scheme as in our fiducial model and use the same
feedback parameters as in Guo11, except for the reheating efficiency
$\epsilon_{\mathrm{reheat}}$, which is slightly reduced in order to
match the observed stellar metallicity content of galaxies in the
local Universe.  

\subsubsection{Halo-mass dependent scaling for the gas
  re-incorporation (Hen13)}  
%******************************************************************************

\citet{Henriques13} use a Monte Carlo Markov Chain approach to explore 
the parameter space of the \citet{Guo11} model, focusing on matching
the observed evolution of the GSMF. They find that no simple
modification of the parameter set is able to reproduce the observed
galaxy number densities from $z=0$ to $z\sim3$. The suggested solution 
is a modification of the timescale for gas re-incorporation that
scales with halo mass. Specifically, they use the same parametrization
for the reheating and ejection rate adopted in \citet{Guo11} with
different parameters:  
\begin{equation}
\dot{M}_{\mathrm{reheat}} = \epsilon_{\mathrm{reheat}} \left[ 0.5 +
\left( \frac{V_{\mathrm{max}}}{336\ \mathrm{km/s}} \right)^{-0.46} \right]
\times \dot{M}_{\mathrm{star}},
\end{equation}
\begin{equation}
\dot{E}_{\mathrm{FB}} = \epsilon_{\mathrm{eject}} \left[ 0.5 +
\left( \frac{V_{\mathrm{max}}}{405\ \mathrm{km/s}}
\right)^{-0.92}\right] \times 0.5  \dot{M}_{\mathrm{star}}
V_{\mathrm{SN}}^2.
\end{equation}
The timescale for gas re-incorporation is assumed to depend on the
inverse of halo mass:
\begin{equation}\label{reinc_eq}
{M}_{\mathrm{reinc}} = \gamma
\frac{{M}_{\mathrm{eject}}}{t_{\mathrm{reinc}}},\ \mathrm{with}\
t_{\mathrm{reinc}} = \frac{10^{10}M_\odot}{M_{\mathrm{vir}}} \times yr.
\end{equation}
This model reproduces the observed evolution of the stellar mass and luminosity
functions from $z=0$ to $z\sim 3$, by construction.

\subsubsection{Early stellar feedback (zDEP)}  
%******************************************************************************

\citet{Stinson13} and \citet{Kannan14} introduced the so called
``early stellar feedback'', likely originating from radiation pressure
from young massive stars at high redshifts. Their cosmological
simulations, carried out using the Gasoline code and including a SN
feedback modelled in the form of `blast-waves', were shown to
successfully reproduce the cosmic evolution of baryon conversion
efficiencies in low-mass halos. 

Inspired by the success of these sub-grid models and in an attempt to
mimic their effect, we introduce an `ad-hoc' model, where we assume
feedback efficiencies to depend on redshift (higher reheating and
ejection rates are assumed at higher redshifts). Starting from the
Guo11 feedback scheme, we implement the following parametrizations, by
adding a dependence on $(1+z)^3$ for both the ejection and re-heating
efficiencies: 
\begin{equation}
\dot{M}_{\mathrm{reheat}} = \epsilon_{\mathrm{reheat}} \left[ 0.5 +
  (1+z)^3 \left( \frac{V_{\mathrm{max}}}{70\ \mathrm{km/s}}
  \right)^{-3.5} \right] \dot{M}_{\mathrm{star}},
\end{equation}
and
\begin{equation}
\dot{E}_{\mathrm{FB}} = \epsilon_{\mathrm{eject}} \left[ 0.5 + (1+z)^3 
\left( \frac{V_{\mathrm{max}}}{70\ \mathrm{km/s}} \right)^{-3.5}\right]
 0.5  \dot{M}_{\mathrm{star}} V_{\mathrm{SN}}^2.
\end{equation}
In this scheme, the reincorporation of gas is modelled as in the
fiducial model. We stress that, although the proposed scenario is
physically motivated, the adopted parametrizations are `empirical',
and obtained by forcing the model to reproduce the observed evolution
of the low-mass end of the GSMF.

\subsubsection{Stellar feedback parameterized from cosmological zoom 
  simulations (FIRE)}  
%******************************************************************************

\citet{Hopkins14} developed a sophisticated sub-resolution model
accounting for  individual sources of stellar feedback in the form of
energy and momentum input from SN explosions, radiative feedback
(photo-heating and radiation pressure), and stellar winds (`the
Feedback In Realistic Environments' - FIRE). Performing fully
cosmological simulations, the authors showed that these models are
successful in reproducing the cosmic evolution of baryon conversion
efficiencies in low-mass halos. Their simulations reveal that the
high-redshift galaxy evolution is dominated by strong bursts of star
formation, followed by powerful and highly non-linear gusts of
galactic outflows that sweep up large fractions of the gas of the
interstellar and circum-galactic medium of galaxies. At
low redshift, sufficiently massive galaxies switch into a continuous
and quiescent mode of star formation that does not drive outflows. In
a recent work, \citet{Muratov15} have parametrized the outflow rates
of gas in these simulations, finding an explicit dependence on
redshift at a given circular velocity (see their eqs. 4 and 5), or
equivalently, a dependence on stellar mass (see their eq. 8). 

Adopting the parametrizations given in \citet{Muratov15}, we assume:  
\begin{equation}
\dot{M}_{\mathrm{reheat}} = \epsilon_{\mathrm{reheat}} (1+z)^{1.25}
\left( \frac{V_{\mathrm{max}}}{60\ \mathrm{km/s}} \right)^{\alpha} 
\times \dot{M}_{\mathrm{star}}.
\end{equation}
For circular velocities $V_{\mathrm{max}} < 60\ \mathrm{km/s}$,
the exponent $\alpha$ is found to be  $-3.2$, while for circular
velocities $V_{\mathrm{max}} > 60\ \mathrm{km/s}$, $\alpha$ has a
value  of  $-1.0$. This means that the FIRE simulations predict
``momentum-driven'' winds for massive galaxies, but ``stronger'' winds      
for less massive ones. 

We then assume: 
\begin{equation}
\dot{E}_{\mathrm{FB}} = \epsilon_{\mathrm{eject}} (1+z)^{1.25}
\left( \frac{V_{\mathrm{max}}}{60\ \mathrm{km/s}} \right)^{\alpha}
\times 0.5 \dot{M}_{\mathrm{star}} V_{\mathrm{SN}}^2,  
\end{equation}
and, as in the Guo11 model, we calculate the ejected gas mass from the
hot gas reservoir following energy conservation arguments
(Eq. \ref{Encons}). We have verified that the alternative
parametrization given for gas outflows in \citet{Muratov15}, including
an explicit dependence on stellar mass, leads to results very similar
to those obtained with the above parametrization. 

 \citet{Muratov15} have not quantified the time-scales over which gas
 is  re-incorporated back onto the halo. We find that, in this
 feedback model, gas  re-incorporation needs to  be delayed as
 proposed in the Hen13 scheme  in order to match the present-day  GSMF
 (particularly around the  exponential cut-off). 

\subsubsection{Preventive feedback (PREH)} 
%******************************************************************************

The last feedback scheme we have implemented is based on what we call  
a `preventive feedback' model.  The circum-galactic/intergalactic
medium may be preheated (to some level of entropy) by early feedback
processes, such that the amount of infalling (for the first time)
pristine gas is reduced in lower mass halos with respect to the
universal baryon fraction. Physical mechanisms for pre-heating are not
well constrained: they are likely related to various phenomena
including e.g. stellar/AGN-driven winds (\citealp{Mo02, Mo04}) or
intergalactic turbulence (\citealp{Zhu11}). Inspired by the work by
\citet{Lu07} and \citet{Lu15}, we have tested the following scheme: we
reduce the amount of infalling gas by a factor $f_{\mathrm{preheat}}$,
linearly scaling with the halo virial mass $M_{\mathrm{vir}}$ for
galaxies with halo masses below $M_{\mathrm{crit}} = 10^{12} M_\odot$
at all redshifts: 
\begin{equation}
f_{\mathrm{preheat}} = \frac{M_{\mathrm{vir}}}{M_{\mathrm{crit}}}
\end{equation}
The rate of newly infalling gas $\dot{M}_{\mathrm{gas, infall}}$ is
then given by: 
\begin{equation}
\dot{M}_{\mathrm{gas, infall}} = f_{\mathrm{preheat}} \times
f_{\mathrm{bar}} \times \dot{M}_{\mathrm{DM, infall}},
\end{equation}
where $f_{\mathrm{bar}}$ is the universal baryon fraction and
$\dot{M}_{\mathrm{DM, infall}}$ the dark matter accretion rate.  For
estimating  $f_{\mathrm{preheat}}$, we have not added any additional
redshift dependence or more complex dependence on halo mass (as in
\citealp{Lu15}). 

For simplicity, we add the amount pre-heated gas to the ejected gas 
component. Therefore, this gas will be accreted onto the halo at later
times, following the re-incorporation model adopted in Hen13 (equation 
\ref{reinc_eq}). Finally, in this scheme, we model the heating and
ejection due to stellar feedback as in our fiducial scheme.

%******************************************************************************
%******************************************************************************
\section{The `anti-hierarchical' evolution of galaxies}\label{downsizing} 
%******************************************************************************
%******************************************************************************

In this section, we focus on a few basic predictions of our {\sc GAEA} model,
and illustrate to what extent they are affected by different parametrizations
for stellar feedback and gas recycling. The trends discussed in this section,
 i.e. the evolution of the GSMF and of the luminosity function, are often
referred to as `anti-hierarchical', suggesting conflicts with expectations from
the currently favoured cosmological model for structure formation.

\subsection{The evolution of the galaxy stellar mass function}

\begin{figure*}
  \centering
  \epsfig{file=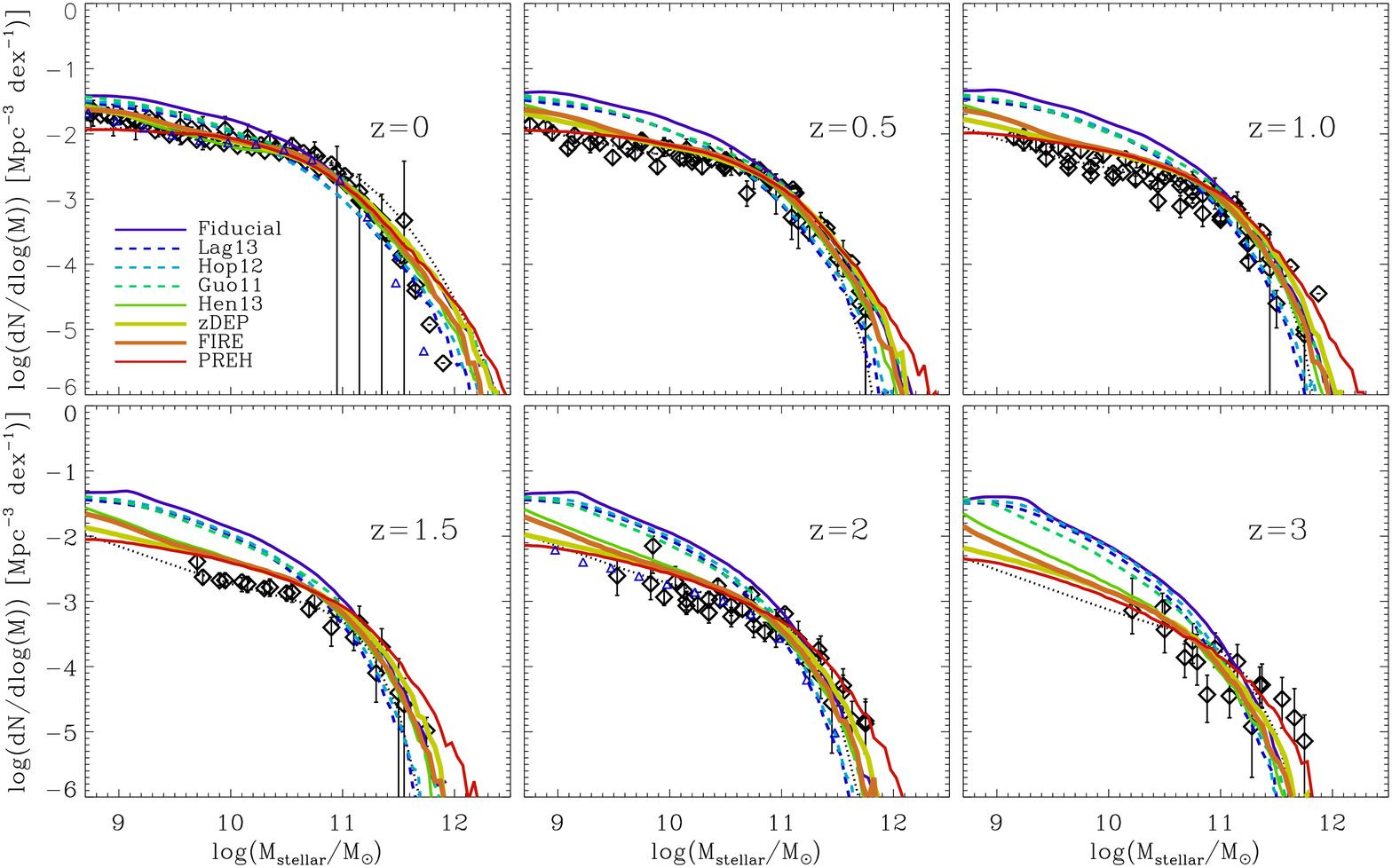, width=0.8\textwidth}
  \caption{Redshift evolution of the GSMF for the different feedback models
    considered in this study (lila lines: fiducial, dark blue lines: Lag13,
    light blue lines: Hop12, turquoise lines: Guo11, green lines: Hen13,
    yellow: zDEP, orange: FIRE, red: PREH), compared to observational
    measurements (\citealp{Bell03, Perez08, Bundy05, Drory04, Fontana06,
      Panter07, Marchesini08, Ilbert10, Muzzin13}, black symbols and black
    dotted lines). Both strong ejective feedback models (Hen13, zDEP,
      FIRE) and the preventive feedback scheme (PREH) are successful in
      reproducing the measured evolution of the low-mass end of the GSMF.}
          {\label{SMF_evol}}
\end{figure*}

Fig. \ref{SMF_evol} shows the evolution of the GSMF for the different
models used in this study (lines with different colours) from $z=3$ to
present (different panels), compared to various observational
measurements (black symbols and dotted lines). Observational data
clearly illustrate the `anti-hierarchical' trend mentioned above: the
massive end appears to be in place already at high redshifts
($z=2-3$), while the number densities of low-mass galaxies increase
significantly towards $z=0$. 

The fiducial model strongly over-estimates the number density of low-mass
galaxies, particularly at high redshifts, i.e. too many low-mass galaxies form
at early times. This has been a well-known problem for galaxy formation models
for at least one decade now \citep[e.g.][]{Fontana04, Fontanot09,
  Weinmann12}. Increasing the feedback efficiency does not improve the
  agreement with observational measurements in this model because it reduces
  the number density of galaxies around the knee of the mass function.

We find the same qualitative behaviour for the Lag13, Hop12 and Guo11 
models, although these are in better agreement (by construction) with
the observational measurements at $z=0$. This means that the different
dependencies on other physical quantities such as gas surface density,
cold gas fraction or circular velocity, at least in the form of the
patametrizations considered here, are not adequate to efficiently
decouple the evolution of baryons from that of low-mass dark matter
haloes, in particular at early times.  

Assuming longer time-scales for gas re-cycling (Hen13, FIRE), a
stronger energy input at early times, i.e. an explicit redshift
dependence of the ejected and reheated gas (zDEP, FIRE), or some form
of pre-heating (PREH), can significantly reduce the number density of
low-mass galaxies ($<10^{11} M_\odot$) by up to one order of
magnitude, particularly at high redshifts. All these modifications are
able to bring model predictions in fairly good agreement with the
observed evolution of the GSMF over the redshift range considered.

We stress that a crucial element for the success of the feedback
schemes Hen13, zDEP and FIRE is that large fractions of gas (larger
than the amounts reheated by stellar feedback) can be ejected outside
the haloes. When assuming a redshift dependent ejection rate and/or
longer gas re-cycling timescales in the framework of the fiducial
feedback scheme, where the amount of gas that can be ejected is
limited to that reheated (according to the energy driven formulation
adopted), we do not reproduce the evolution of the GSMF. 

 Our results indicate that both a strong ejective (Hen13,
  zDEP, FIRE) and a preventive form (PREH) of  feedback are
  capable of reproducing the observed evolution of the GSMF. To what
extent these models are able to reproduce other galaxy 
physical properties, and how we can discriminate which scheme performs
better, will be discussed below in sections \ref{Gas} and \ref{Stellarpop}. 

\subsection{The evolution of the galaxy luminosity functions}

\begin{figure*}
  \centering
  \epsfig{file=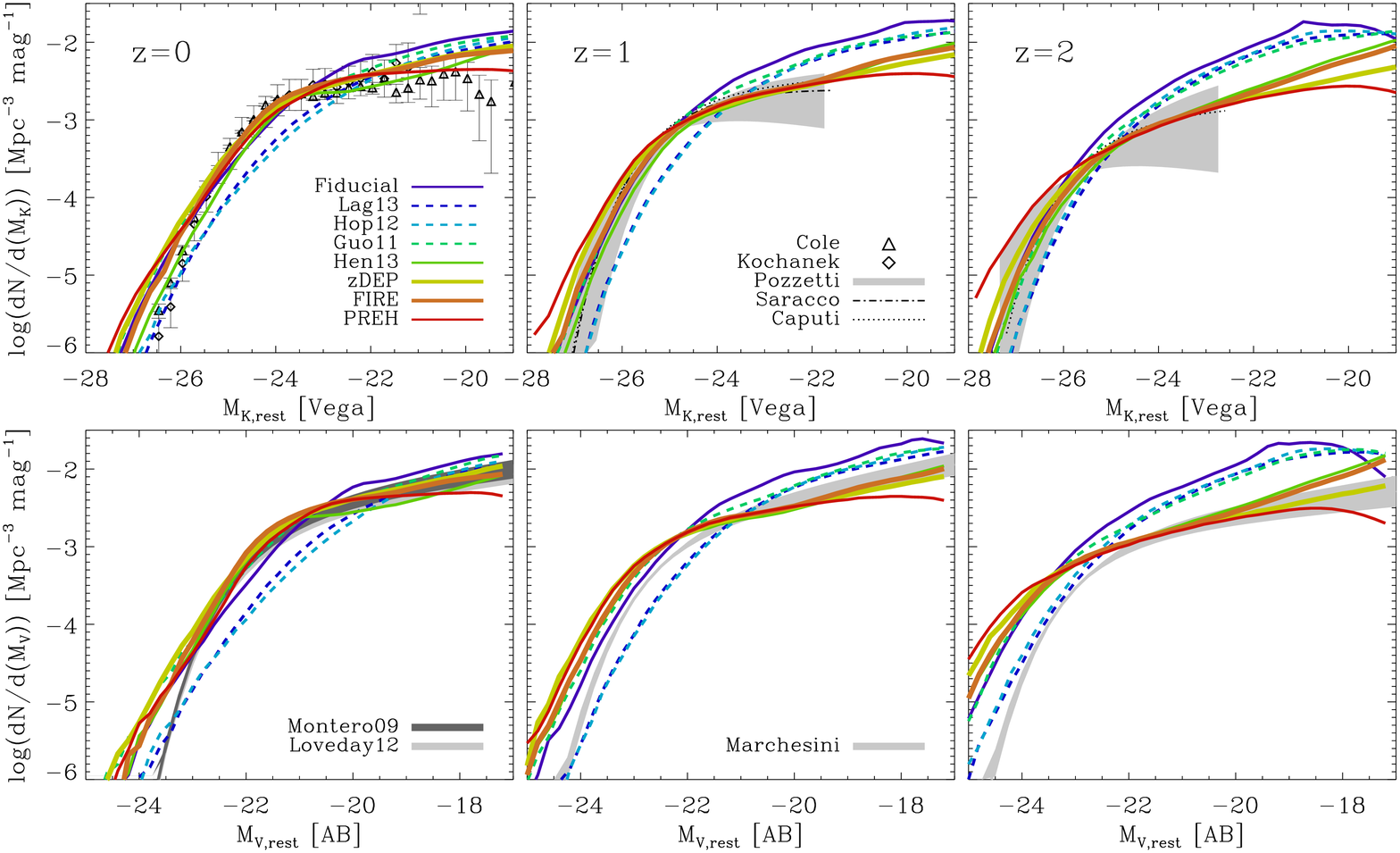, width=0.8\textwidth}
  \caption{Evolution of the K-band (top row) and V-band luminosity function
    (bottom row) in different models (colored lines as in Fig. \ref{SMF_evol}), 
compared to observational measurements by \citet{Cole01, 
      Kochanek01, Pozzetti03,  Saracco06, Caputi07} and
    \citet{Marchesini12} (black symbols and grey shaded areas).}  
{\label{Lumfct}} 
\end{figure*}

It is well known that the conversion from observable properties (typically,
luminosities in different bands) to stellar masses carries relatively large
uncertainties, both statistical and (more dangerous because less under control)
systematics (e.g. see discussion in \citealt{Marchesini09} and
\citealt*{DeLucia14b}). On the other hand, galaxy formation models can predict
observables, specifically luminosities, albeit making a number of assumptions
(the same needed to convert observed luminosities into physical
  properties) on e.g. the stellar population models, the initial mass
function, dust attenuation. 

In Fig. \ref{Lumfct}, we show the K-band (top panels) and
V-band (bottom panels) galaxy luminosity functions at $z=0, 1, 2$ from
the models (lines of different colours), compared to observational
measurements by \citet{Cole01, Kochanek01, Pozzetti03, Saracco06,
  Caputi07} and \citet{Marchesini12} (black dotted and dashed-dotted
lines and grey shaded areas). Luminosities are computed assuming a
Chabrier initial mass function (consistently with the choice adopted
for the chemical enrichment) and stellar population synthesis models
by \citet{Bruzual03}, as described in \citet{DeLucia04}. 

At $z=0$, the strong ejective feedback models (zDEP, FIRE, and Hen15)
still slightly over-estimate the number densities at the low-luminous
end of the K-band luminosity function. The PREH model appears to
provide a better match to the observational data. As for the galaxy
stellar mass function, the Hen13, zDEP, FIRE and PREH models predict a
relatively shallow faint end slope, in fairly good agreement with
observations at $z=1$ and $z=2$. 

For the V-band luminosity function (measured down to $M=-17$ at
$z=2$), the PREH model even tends to slightly under-estimate the
number densities of low-luminous galaxies at $z=1 - 2$, while the
Hen13 and FIRE models still slightly over-predict the number densities
of low-luminosity galaxies at $z=2$. The zDEP model provides the best 
match to the observed V-band luminosity function over the redshift
range considered. 

Turning to the bright end of the luminosity function, we find an
underestimation of luminous galaxies in the Hop12 and Lag13 models,
irrespective of the redshift. This is relevant around L$_*$. Only at the
highest redshift considered ($z\sim2$), these models predict number
densities for the most luminous galaxies that are in agreement with
observational determinations. All other models tend instead to
over-estimate the bright end, particularly for the V-band luminosity
function. These over-abundant V-band luminous galaxies in the models
have typically experienced recent merger driven starbursts. It is
likely that our treatment of dust attenuation is inadequate for these
systems, which might explain the relatively good agreement with the
observed number densities of the most luminous galaxies in the K-band,
and the poorer agreement for the V-band.

\subsection{The evolution of the baryon conversion efficiencies}

\begin{figure*}
  \centering
  \epsfig{file=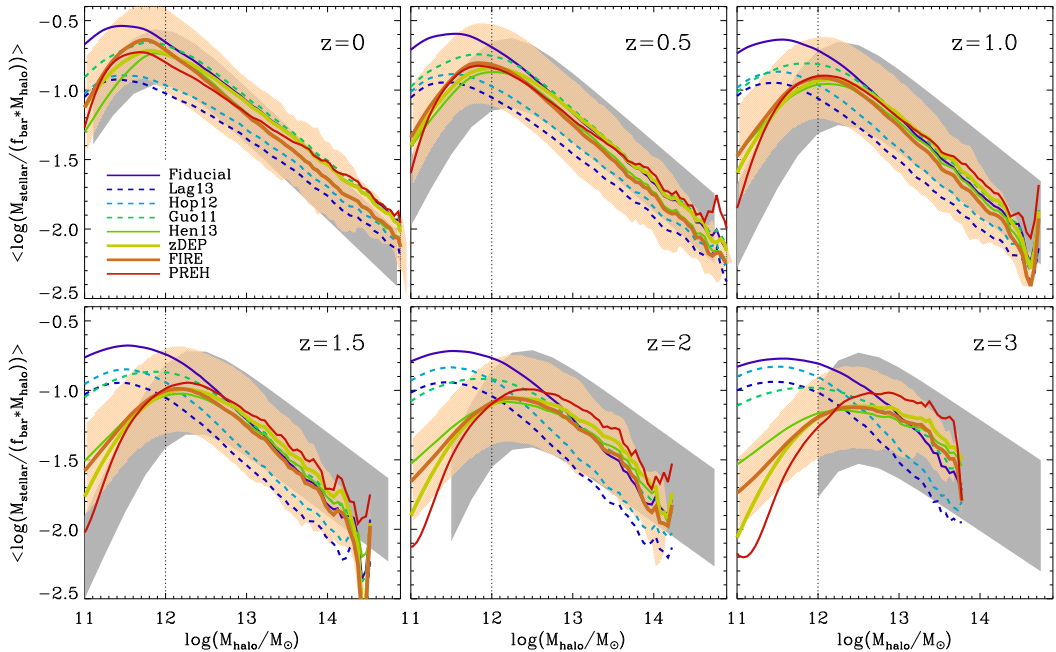, width=0.8\textwidth}
  \caption{Mean Baryon conversion efficiencies
    ($M_{\mathrm{stellar}} /  (f_{\mathrm{bar}} \times
    M_{\mathrm{halo}})$) versus halo mass at different redshifts, as
    predicted by the different models used in this study 
    (colored lines as in Fig. \ref{SMF_evol}; the orange dashed area
    illustrates the 1-$\sigma$-scatter corresponding to the FIRE
    model). The Hen13, zDEP, FIRE and PREH models  (green,
      yellow, orange, red solid lines), that match the GSMF, are also 
    consistent with predictions from the subhalo abundance 
    matching method by \citet{Moster13} (grey shaded areas). The
    vertical dotted lines correspond to halo mass of $\sim 10^{12}
    M_\odot$. At the resolution of the Millennium, these are resolved
    with less than 1000 particles. }  {\label{Barconv}}   
\end{figure*}

Fig. \ref{Barconv} shows the redshift evolution of the baryon
conversion efficiency versus the dark matter halo mass for the
different feedback models used in this study (lines of different
colours), compared to predictions from the subhalo abundance 
matching  approach discussed in \citet{Moster13} (grey shaded
areas, see also \citealp{Behroozi13}). The baryon conversion
efficiency is defined as the ratio between the total stellar mass
$M_{\mathrm{stellar}}$ within a given halo and the total baryonic mass
expected assuming a  universal baryon fraction ($f_{\mathrm{bar}}
\times M_{\mathrm{halo}}$ - we assume here $f_{\mathrm{bar}} =
0.18$). At the mass resolution of the Millennium simulation, haloes
with mass $\sim 10^{12}\,{\rm M}_{\odot}$ (marked by vertical
dotted lines in Fig. \ref{Barconv}) are resolved with less than
1000 particles. For the figure shown, we are therefore pushing into 
the resolution limit of our simulation. We have, however, verified
that results based on the Millennium II are consistent with those
shown. These convergence tests are discussed in Appendix
\ref{resolution}. 

The subhalo abundance matching method gives a maximum baryon
conversion efficiency of 20-30 per cent, for halo masses of roughly
$10^{12} M_\odot$. For higher and lower masses, the conversion
efficiencies drop to much smaller values, down to only a few per
cent. In addition, the peak of the baryon conversion efficiency is
found to shift towards lower halo masses with decreasing redshift. In
other words, the halo mass scale at which most of the star formation
takes places increases at higher redshift.

Irrespective of the redshift, the fiducial, Lag13, Hop12 and Guo11
models predict by far too high conversion efficiencies in halos less
massive than $10^{12} M_\odot$. In addition, they cannot capture the
correct time evolution of the peak: the fiducial, Lag13 and Hop12
models predict a maximum conversion efficiency at halo masses of $\sim
10^{11.5} M_\odot$, while the Guo11 model predicts a maximum
conversion efficiency at $\sim 10^{12} M_\odot$. The peak hardly
evolves with redshift in all these models. The Hen13, zDEP, FIRE and
PREH models predict baryon conversion efficiencies in low mass haloes
that are in good agreement with expectations, and the correct
evolution as a function of redshift. This confirms results from the
previous sections: both strong ejective (Hen13, zDEP, FIRE) and
preventive (PREH) feedback models can capture the observed trends. On  
the basis of the results presented so far, however, we cannot
discriminate among the proposed solutions.  

%******************************************************************************
%******************************************************************************
\section{Origin of the `anti-hierarchical' trend}\label{origin}  
%******************************************************************************
%******************************************************************************

In order to understand the origin of the `anti-hierarchical' trends
illustrated in the previous section, and the physical mechanisms
responsible for the success or failure of the proposed feedback
schemes, we analyse the circulation of baryons between the different
phases (i.e. the amount of gas heated, ejected, re-incorporated or
cooled and being converted into stars), for the different
parametrizations adopted in this study.

\begin{figure*}
  \centering
    \epsfig{file=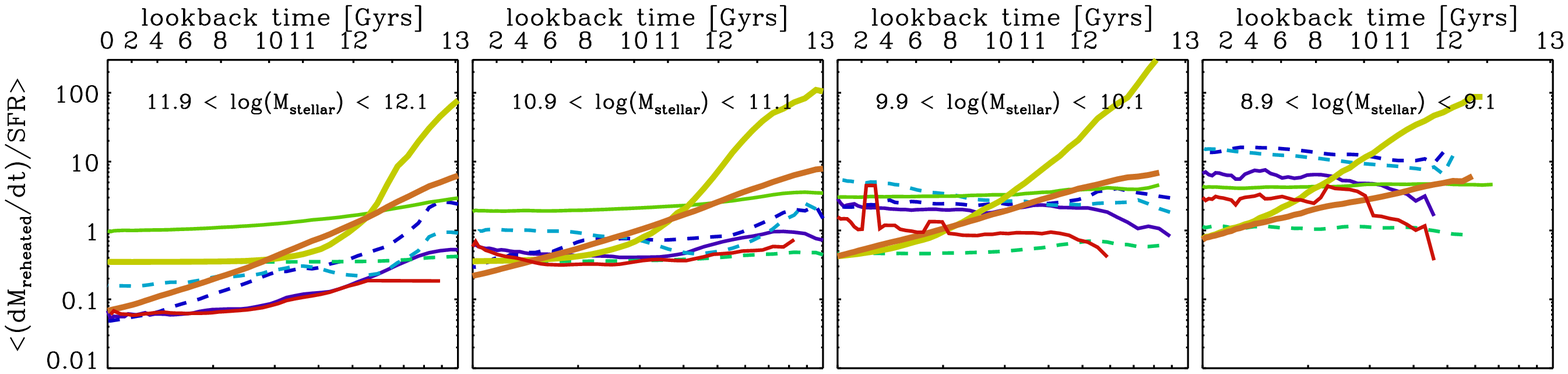,
      width=0.95\textwidth}\vspace{-1.5cm} 
    \epsfig{file=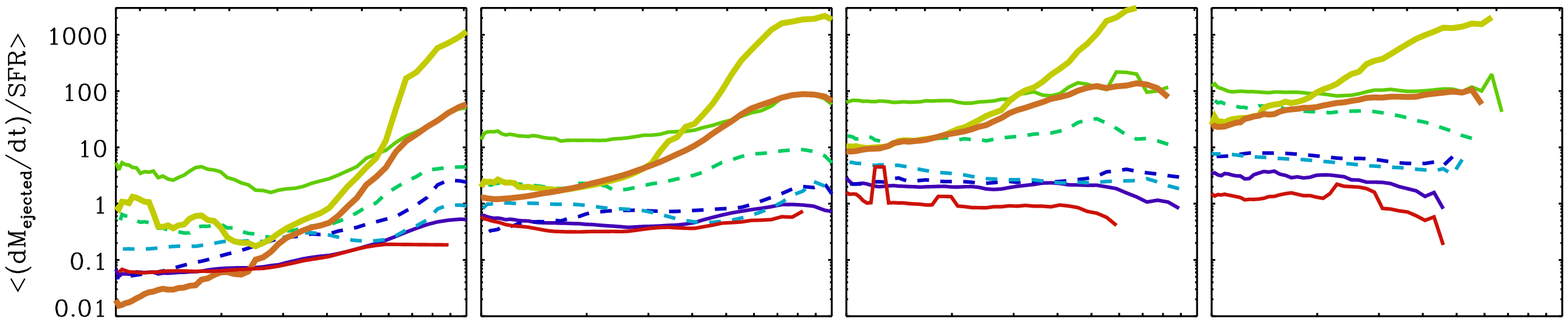,
      width=0.95\textwidth}\vspace{-1.5cm} 
     \epsfig{file=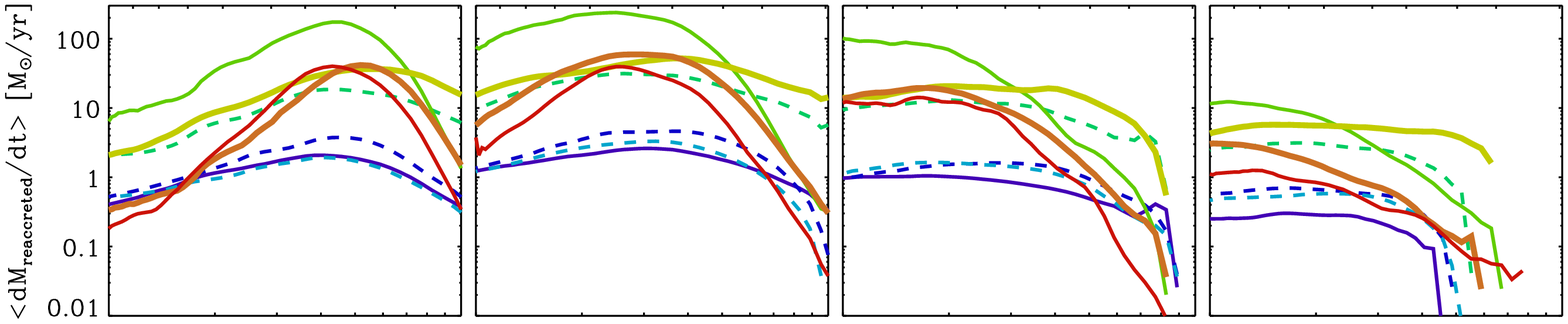, width=0.95\textwidth}\vspace{-1.5cm}
  \epsfig{file=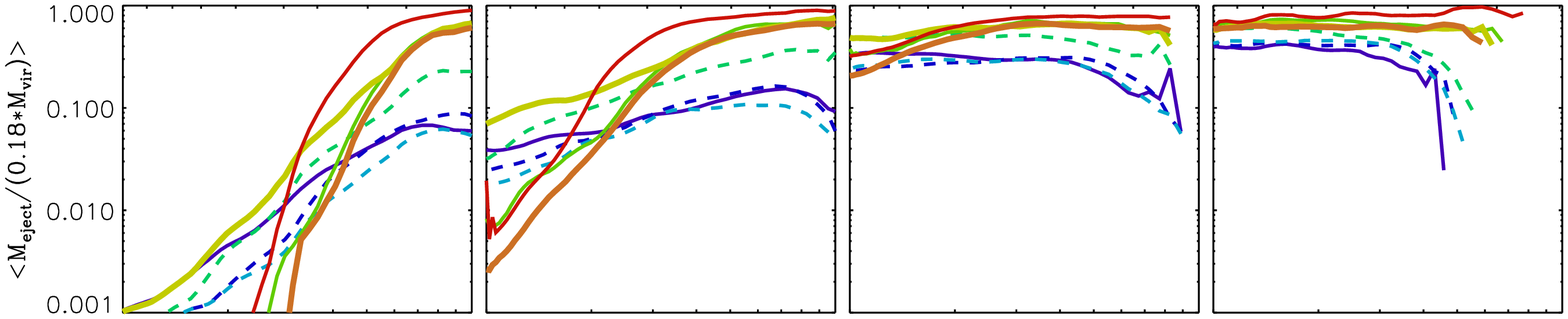, width=0.95\textwidth}\vspace{-1.5cm}
  \epsfig{file=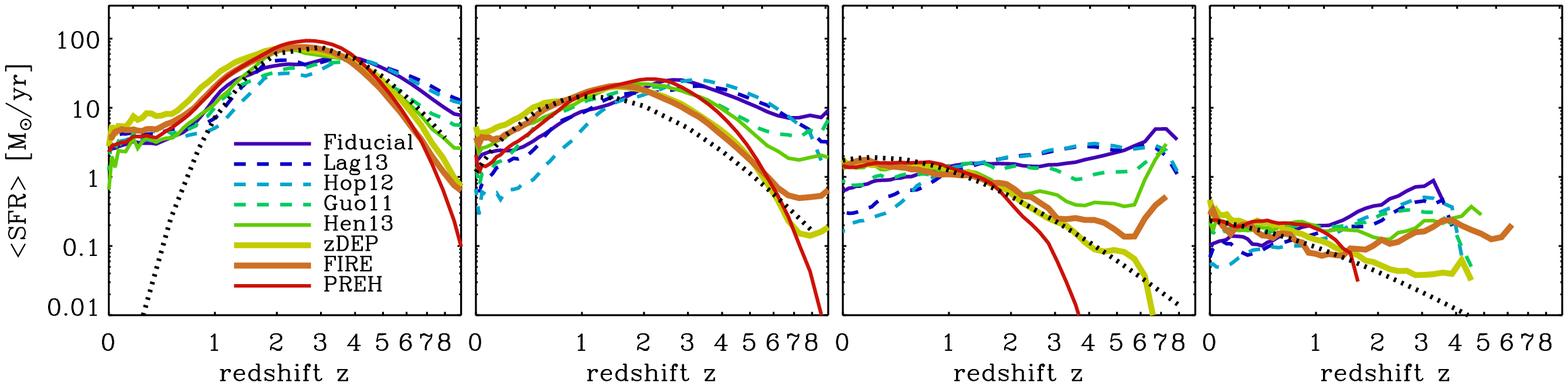, width=0.95\textwidth}
  \caption{Redshift evolution of the  mean mass-loading for
    the reheated (first row), ejected (second row), and re-accreted
    rate (third row). The fourth row shows the evolution of the
      mean fraction of baryons in the ejected gas phase and the
    bottom row the evolution of the mean  SFRs. Different
    columns correspond to  galaxies of different present day stellar
    mass. Lines of different colours (as in  Fig. \ref{SMF_evol})
    illustrate the different  feedback schemes  considered in this
    work. Black dotted lines in the bottom  row show estimates based
    on subhalo abundance matching methods by \citet{Moster13}.}
  {\label{Baryon_evol}} 
\end{figure*}

\subsection{Evolution of the gas outflows}

The first and second row of Fig. \ref{Baryon_evol} show the redshift
evolution of the mean reheated and ejected gas mass rates (due
to stellar feedback) for the different feedback schemes tested in this
work (illustrated by lines of different colours). Results are shown
for galaxies in four different bins of stellar mass (different
columns). Since ejection and re-heating rates scale linearly with
  the SFRs (see equations summarized in the second and third column in
  table 1), we have normalised them to these values so that the
figures show what are usually referred to as `mass-loading'
factors. To compute the averages shown in the figure, we have selected
galaxies on the basis of their galaxy stellar mass at $z=0$, and
traced their main progenitor (typically the most massive progenitor at
each node of the merger tree) backwards in time.  

In the strong ejective feedback models (Hen13, zDEP and FIRE),
the mass-loading of the re-heated gas is by up to two orders of
magnitude larger than in the fiducial, Lag13, Hop12, Guo11, and PREH
models, particularly at redshift $z>2$. As noted above, our
  preventive feedback scheme (PREH) adopts the same modelling for the
reheating and ejection as our fiducial model. Therefore, the predicted
evolution are very similar but both the reheating and the ejection
rates are systematically lower in the former scheme, at all redshifts
for low mass galaxies, and at high redshift for more massive systems:
less gas is available for cooling at high redshift in small haloes, so less
stars are formed and less gas can be reheated.

When looking at the ejection rate of gas, the differences between the
feedback schemes are even more extreme. For the fiducial, Lag13, Hop12
and the PREH model, the (ejection) mass loading is rather low and
hardly evolves with redshift. Instead, for the Guo11, Hen13, zDEP and
FIRE models, the mass-loading can be by up to three orders of
magnitude larger. This is a direct consequence of the higher reheating
rates, and is enhanced by the assumption that also a fraction of the
hot gas associated with the parent halo (and not only the gas reheated
from the cold disk component) can be driven out of the halo.
 
The FIRE and zDEP models are characterised by a very strong decrease
of the mass-loading factor with time, resulting in significantly lower
ejection rates at $z=0$ than e.g. in the Hen13 model. In this model,
the mass-loading is in fact high, but nearly constant as a function of
cosmic time. We will come back to consequences of these different gas
flows on e.g. the metal enrichment of model galaxies in section
\ref{Stellarpop}. The results discussed so far indicate that high
ejection rates at early times represent a possible (but not necessary)
condition for reproducing the observed evolution of the GSFM. 

\subsection{Evolution of the gas re-accretion rate}

The third row of Fig. \ref{Baryon_evol} shows how the re-accretion gas rates
(these are, in all models, proportional to the amount of gas in the
  ejected reservoir) vary as a function of redshift. As a direct consequence
of the larger amount of ejected material in the Guo11, Hen13, zDEP and FIRE
models, the mean gas re-incorporation rates are larger than in the fiducial,
Lag13 and Hop12 models, but at low redshift for the most massive galaxies
considered. Here, the re-accretion rates predicted by the FIRE model are
comparable to, or even lower than those found in the fiducial scheme.

The high average re-accretion rates predicted by the PREH model are a
result of our specific treatment for the gas that is prevented to fall
onto dark matter haloes.  For simplicity, we assign this gas to the
same reservoir as the ejected component (both are not available for
cooling). This obviously leads to high `re-accretion' rates although
the term is inappropriate in this case since most of these baryons
have never been inside the halo before. 

The figure also shows the impact of the adopted gas recycling scheme:
models assuming re-incorporation time scales inversely proportional to
halo mass (Hen13, FIRE, and PREH) predict a different redshift
dependence of the gas re-accretion rates than the other models
assuming the gas to be re-incorporated over a halo dynamical time
scale. Independently of the galaxy stellar mass considered, a halo
mass dependence of the re-incorporation time scales leads to a strong
suppression of early gas re-accretion rates with respect to the
fiducial reincorporation scheme. Gas re-accretion is delayed to
progressively lower redshifts for galaxies of decreasing stellar mass,
resembling the `anti-hierarchical' trends discussed earlier. In
contrast, models based on the fiducial gas re-incorporation scheme do
not reproduce such a behaviour, and predict nearly constant
re-incorporation rates as a function of redshift.

%% As discussed in \citet{Henriques13}, the suppression of gas
%% re-incorporation at early times prevents gas cooling, and thus, delays
%% star formation towards later times (in particular in low-mass
%% galaxies), clearly helping to reproduce the observed trends. In fact,
%% our results demonstrate that these rather originate by the (complex)
%% interplay between gas reheating, ejection and re-incorporation,
%% i.e. all the different components of the `feedback' scheme. 

\subsection{Evolution of the ejected baryon fraction}

The results shown in the previous sections demonstrate that the model
that successfully reproduce the measured evolution of the GSMF (Hen13,
zDEP, FIRE, and PREH) are characterised by rather different behaviours
of the ejection and re-incorporation rates.

One crucial, common feature of these models is the existence of large
amounts of baryons in the ejected component, where they are not
available to cooling. The fourth row of Fig. \ref{Baryon_evol} shows
the evolution of the mean `baryon ejected fractions' as a
function of redshift. These are defined as the fractions of baryons
residing in the ejected component with respect to the expected
baryonic mass ($=M_{\mathrm{eject}} / (f_{\mathrm{bar}} *
M_{\mathrm{halo}})$). We stress that for the PREH model, the ejected
component comprises (by construction) gas prevented from infall,
i.e. that was never within a halo before. 

Hen13, zDEP, FIRE, and PREH require the vast majority of baryons/gas
to reside in the `ejected' reservoir. For low mass galaxies ($10^9
M_\odot$), the ejected fractions are always very high (larger than 60
per cent). For more massive galaxies, the fraction of baryons in the
ejected component is very large at high redshift and decreases with
decreasing redshift, more rapidly so for more massive galaxies. For
the most massive galaxies considered, the ejected fractions are lower
than those predicted by the fiducial model for $z<2-3$ in the Hen13,
FIRE and PREH models. 

Therefore, to falsify predictions from both successful ejective
  (zDEP, FIRE) and preventive (PREH) feedback models, observations of
the (diffuse, weakly ionised or neutral) gas in the circum-galactic
(CGM) and/or inter-galactic medium (IGM) are essential. Diffuse warm
gas ($10^4 - 10^5$ K) is very difficult to detect, but high-resolution
spectroscopy in the rest-frame UV has started to probe the diffuse gas
and metals in the CGM using absorption line measurements along the
line of sight of quasars (e.g. \citealp{Peeples14, Prochaska13, Rudie12,
Tumlinson13}). This provides more stringent constraints on the gas and
metals that have been ejected by the winds or have been prevented from
infall due to pre-heating invoked by our models.  In case of
preventive feedback, for example, the CGM/IGM should mainly consist of
pristine gas, while for ejective feedback schemes, we would expect
much higher levels of metal enrichment.

\subsection{Evolution of star formation rates}

As a direct consequence of the large ejected gas fractions in the Hen13, zDEP,
FIRE, and PREH models, star formation is suppressed at early times for galaxies
of all masses. The bottom row of Fig. \ref{Baryon_evol} shows the mean
  star formation histories of galaxies in each of the stellar mass bins
considered, and compare predictions from the feedback schemes analysed in this
study with estimates based on the subhalo abundance matching method by
\citet{Moster13}. For galaxies less massive than $10^{11}\,{\rm M}_{\odot}$,
the SFRs in the models are \textit{increasing} with decreasing redshift, while
the opposite behaviour is predicted by the fiducial, Lag13, Hop12, and Guo11
models. For the most massive galaxies, the SFRs peak at higher redshift than
for their lower mass counterparts in the Hen13, zDEP, FIRE and PREH models,
while such a trend is weaker or insignificant for the other models.

The behaviour described above is in fairly good agreement with
predictions from the subhalo abundance matching approach described by 
\citet{Moster13}. For low mass\footnote{Note that the $10^9 M_\odot$
  stellar mass bin is at the limit of   the Millennium
  resolution. Using merger trees from the better resolved
  Millennium-II simulation, we have verified that the results are
  unchanged.} galaxies ($10^9-10^{10} M_\odot$), the zDEP model
provides the best match to the estimates by \citet{Moster13}, while
the FIRE and Hen13 models predict larger SFRs at high redshifts. In
the PREH model, the formation of these low-mass galaxies appears to be
simply shifted towards later times. 

For the most massive galaxies ($M_{\mathrm{stellar}} \sim 10^{12}
M_\odot$), the SFRs tend to be over-estimated below $z\sim 1$ with
respect to estimates by \citet{Moster13}, for all feedback schemes
considered in this study. This is due to the fact that, within the new
schemes proposed, the AGN feedback model adopted becomes
inefficient. We will come back to that issue in the following. 

%******************************************************************************
%******************************************************************************
\section{Cold gas in the interstellar medium}\label{Gas}
%******************************************************************************
%******************************************************************************

Due to the large variations of gas reheating, ejection and
re-incorporation rates discussed in the previous section, the stellar 
feedback schemes considered in this study provide significantly
different predictions for the amount and properties of the
interstellar medium in model galaxies. Our current  working version of 
the {\sc Gaea} model follows the \textit{total} cold gas content. An
updated version of our model, that models explicitly the transition
from atomic to molecular hydrogen, is in preparation (Xie et al., in
prep.).

\subsection{Cold gas fractions}\label{Coldgasfrac}

\begin{figure}
\centering
  \epsfig{file=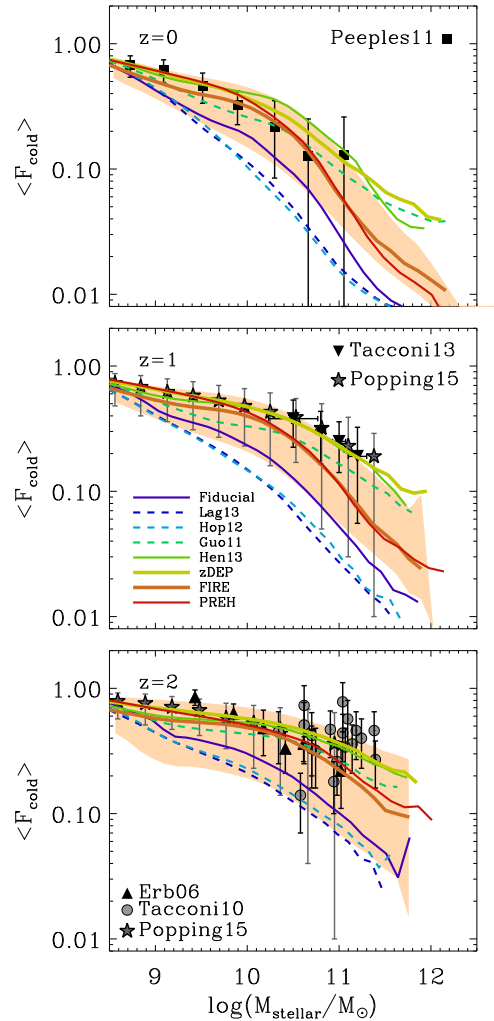,
    width=0.4\textwidth}
 \caption{ Mean cold gas fractions
   ($\langle
   M_{\mathrm{cold}}/(M_{\mathrm{cold}}+M_{\mathrm{stellar}})
   \rangle$) of  star-forming galaxies as a function
   of galaxy stellar mass at z = 0, 1, 2 (from top to bottom). Model
   predictions (colored lines and shaded areas as in
   Fig. \ref{Barconv}) are compared to observational data  (shown as
   black symbols) by \citet{Peeples11, Tacconi13, Tacconi10,
     Popping15} and \citet{Erb06}.}  {\label{fig:Coldgas}}
\end{figure}

Fig. \ref{fig:Coldgas} illustrates how the cold gas fraction varies as
a function of the galaxy stellar mass at z = 0, 1, 2 for the different
feedback models used in this work (lines of different colours), and
compares model predictions to observational data (black and grey
symbols). The orange shaded region illustrates the 1-$\sigma$ scatter
for the FIRE model. Cold gas fractions are defined as:
  
\begin{equation}
F_{\mathrm{cold}} = \frac{M_{\mathrm{cold}}}{M_{\mathrm{cold}} +
  M_{\mathrm{stellar}}},  
\end{equation}
where $M_{\mathrm{cold}}$ is the cold gas mass (in our model this is
associated only with the galaxy disk), and $M_{\mathrm{stellar}}$ is
the galaxy stellar mass.  The model predictions plotted here refer
only to   `star-forming' galaxies, selected according to the criterion 
  suggested by \citet{Franx08}:   $\mathrm{sSFR} >
  0.3/t_{\mathrm{Hubble}}$. In fact, it is worth noting that
  observational measurements are available for relatively small (and
  likely   biased) samples of galaxies, particularly at high
  redshift. In addition, the scatter is rather large ($\pm 0.3$dex),
  in particular for massive galaxies.

At fixed stellar mass, all models predict decreasing cold gas
fractions with decreasing redshift -- at least for galaxies more
massive than $10^{10} M_\odot$. However, only for the Hen13, zDEP,
FIRE, PREH, Guo11 models, gas depletion time-scales are long enough
that the cold gas fractions are consistent with observational
estimates at $z=0, 1, 2$. In contrast, the fiducial, Lag13 and Hop12
models predict gas fractions that are significantly below the
observational estimates, at all galaxy masses, although the scatter
(comparable to that of the FIRE model) is fairly large ($\pm 0.2$
dex). Interestingly, stellar feedback affects strongly the cold gas
content of even very massive galaxies, that are usually believed to be
mainly affected by AGN feedback only. 
 
The cold gas content of a galaxy is determined by the combined effect
of star formation, reheating, and cooling from the hot gas reservoir
associated with the parent halo. The latter quantity depends crucially
on the balance between ejection and re-incorporation. We find that
these two processes are the main responsible for the differences found
among the schemes considered: in the fiducial, Lag13, and Hop12
schemes, the cold gas reservoir of galaxies is quickly consumed via
  star formation at early times ($z>3$), and only little gas is
ejected and re-incorporated at later times. The other models are
characterised by larger ejection rates at high redshift. This prevents
early star formation (that would lock large amount of gas). Later large
re-incorporation rates lead to the large predicted cold gas fractions
that are shown in Fig. \ref{fig:Coldgas}. Again, a very good
  agreement with observed data, down to rather small galaxy stellar
  masses, can be achieved by either ejecting larger fractions than
  reheated gas (i.e. ejecting also a fraction of the hot gas in the
  halo), or preventing gas from infall.

The slightly lower cold gas fractions for galaxies more massive than
$10^{11} M_\odot$ in the FIRE and PREH models with respect to the
Hen13, zDEP and Guo11 models, are due to the lower gas re-accretion
rates, caused by the halo-mass-dependent scaling for gas
recycling. The large gas re-accretion rates (and thus, cold
gas fractions) in the Hen13 model is a consequence of the large and
almost constant ejection rates.

\subsection{Cold gas metallicity}\label{Coldgasmet}

\begin{figure}
  \centering
   \epsfig{file=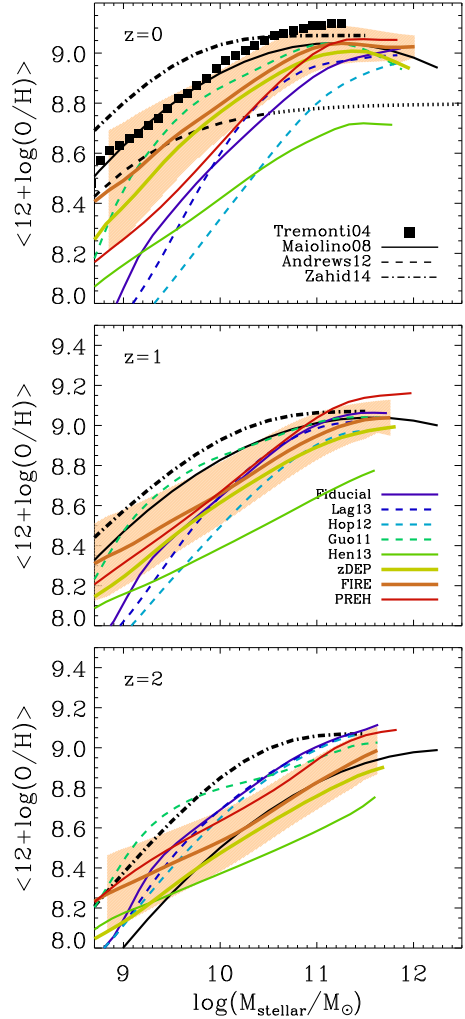,
     width=0.4\textwidth}
 \caption{ Mean cold gas metallicity of star-forming galaxies
   as a function of galaxy stellar mass at z = 0, 1, 2 (from top to
   bottom). Model predictions (colored lines and shaded areas as in
   Fig. \ref{Barconv}) are compared to observational data (shown as
   black lines and symbols) by \citet{Tremonti04, Maiolino08,
     Andrews12}, and \citet{Zahid14}. } 
         {\label{fig:Coldmet}}
\end{figure}

Fig. \ref{fig:Coldmet} shows how the cold gas metallicity of model
  star-forming galaxies (selected as mentioned in the previous
  section) varies as a function of galaxy stellar mass at $z = 0, 1,
  2$ for the different feedback models considered (lines of different
  colours), and compares model predictions with observational
  data. The 1-$\sigma$ scatter is shown only for the FIRE model
  (shaded area), and is representative for all models.  

The absolute value and evolution of the gaseous metallicity provide 
strong constraints on the stellar feedback scheme, due to the
self-consistent modelling of chemical enrichment in our {\sc GAEA}
model. It is worth reminding that the absolute normalisation (and even
the shape) of the observed mass-metallicity relation is strongly
dependent on the choice of metallicity calibration \citep{Kewley08}.
Therefore, it is important that different line diagnostics adopted by
different surveys and/or at different redshifts are cross-calibrated
consistently (e.g. \citealp{Maiolino08}). In this way, the
\textit{relative} evolution of the metal content at fixed stellar mass
is expected to provide a stronger constrain than the absolute value at
a given cosmic epoch.

The PREH, fiducial, Lag13, Hop12, and Guo11 models predict very little 
evolution in the relation between galaxy stellar mass and the
metallicity of the cold gas component, over the redshift range
considered. This is due to the rather low reheating and ejection rates
at early times in these models, that determine high rates of star
formation at high redshift and therefore a rapid enrichment of the
cold gas component of galaxies. 

In the Hen13 model, due to the significantly higher gas re-heating
rates, more enriched material is blown out of the galaxies (and maybe
even out of the haloes) so that the metal enrichment of the cold gas is
delayed towards lower redshift. However, since the high reheating gas
rates are almost constant with time, the metal enrichment remains
strongly suppressed at lower redshifts, leading to an under-estimation
of the cold gas metallicity with respect to observational measurements.

In contrast with the predictions obtained by our Hen13 implementation, 
\citet{Henriques15} argue that the proposed feedback and
re-incorporation scheme predicts a realistic metallicity content for
present-day galaxies. This apparent contradiction is due to the fact
that the model discussed in \citet{Henriques15} assumes an
unrealistically high metal yield (0.047) within their instantaneous
recycling approximation. We will come back to this issue section
\ref{SAMcomp}. 

In the zDEP and FIRE based ejective feedback models, the enrichment of  
the cold gas is delayed at high redshifts as in the Hen13 model, but
the cold gas metallicity is strongly increasing towards $z=0$, in
fairly good agreement with observational constraints from e.g.
\citet{Maiolino08}.  This is due to the rather strongly declining gas
reheating rate with decreasing redshift (see top row in
Fig. \ref{Baryon_evol}) such that much smaller amounts of metal
enriched gas are driven out of the galaxies at low redshifts with
respect to the Hen13 model.

An interesting behaviour is found for galaxies more massive than
$10^{11} M_\odot$: for the zDEP, Guo11 and Hen13 models, the cold gas
metallicity is slightly decreasing with increasing stellar mass -- in
tension with the observed trend. This is most likely a consequence of
the higher re-accretion rates (and thus cooling rates) of relatively
metal-poor gas (that tends to dilute the gas metallicity) with respect
to the other models (see third row in Fig. \ref{Baryon_evol}).

In all models, we find that the slope of the relation hardly changes
with decreasing redshift, while the relation tends to flatten for
galaxies more massive than $M_{\mathrm{stellar}}<10^{10.5}
M_\odot$. This trend is consistent with observational measurements by
\citet{Maiolino08}. Only for the zDEP and FIRE models we find also an
increase in the normalization of the mass-metallicity relation. A more
detailed investigation of the evolution of the mass-metallicity
relation will be the subject of a forthcoming paper. 

%******************************************************************************
%******************************************************************************
\section{Star formation}\label{SFR}
%******************************************************************************
%******************************************************************************

The strong impact of stellar feedback on the amount and composition of
the cold gas content implies rather different star formation histories
in our model galaxies, as discussed in section \ref{origin}. In an
attempt to discriminate among the different schemes analysed in this
paper, we now present a more quantitative comparison between model
predictions and the observed amount of star formation as a function of
stellar mass. 

\subsection{Evolution of the star formation rate function}

\begin{figure*}
  \centering
  \epsfig{file=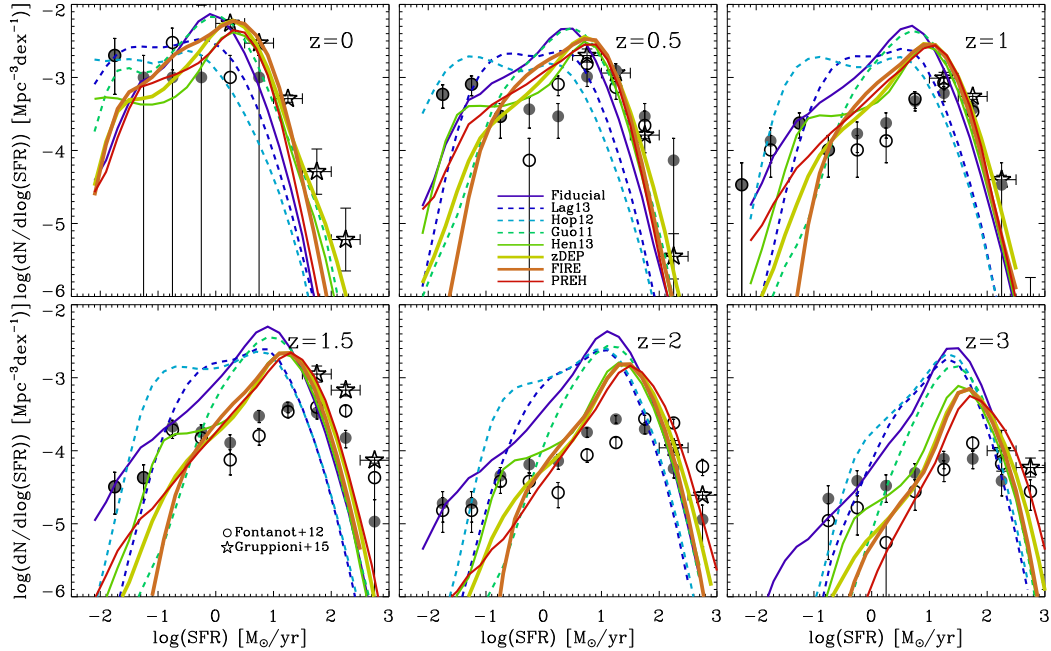,
    width=0.8\textwidth} 
  \caption{Evolution of the SFR function in the different feedback models
   (colored lines as in Fig. \ref{SMF_evol}), compared to
   observational measurements (black circles and stars taken from
   \citealp{Fontanot12, Gruppioni15}; grey circles indicate SFR
   measurements derived from SED fitting only, and are taken from
   \citealp{Fontanot12}). }  {\label{SFRF_evol}} 
\end{figure*}

Fig. \ref{SFRF_evol} shows the evolution of the SFR function for the 
different feedback schemes analysed (lines with different colours),
compared to observational measurements by \citet{Fontanot12}, and
\citet{Gruppioni15} (black circles and stars). \citet{Gruppioni15}
computed the SFRs based on a combination of SED fitting and infrared
{\it Herschel} data coming from the PEP and HerMEs projects (covering
a range from 70 to 500 microns). The SFR functions discussed in
\citet{Fontanot12} are computed from the GOODS-MUSIC catalogues
(\citealp{Santini09}), and they are based either on a combination of
UV and 24 microns (for the high-SFR end of the function) or on SED
fitting (in the low-SFR regime)\footnote{These estimates correspond to
the open circles in Fig. \ref{SFRF_evol}, while filled symbols refer
to the alternative choice of using SED fitting for all GOODS-MUSIC
sources}. The agreement between these different observational data
sets is relatively good, particularly at $z<1.5$.

We adopted a stellar mass cut in the models of $10^{10} M_\odot$ in
order to be consistent with the mass-limited samples of
\citet{Fontanot12}. This mass cut only affects (decreases) the number
densities of galaxies with SFRs below the peak, e.g. SFR $< 1-10
M_\odot/$yr depending on redshift. The comparison with the data by
\citet{Gruppioni15} (IR-flux limited, not mass limited) is, however,
still fair, since the sample includes only galaxies with relatively
high SFRs and the number densities of these is not affected by the
adopted mass cut.

%% The overall SFR functions and their peaks are slightly shifted towards 
%% higher values in the Hen13, zDEP, FIRE and PREH models with respect to  
%% the other models. The former schemes provide a better agreement with
%% data, although the peak of the SFR functions is still too high at most
%% redshift, and it does not follow the measured evolution.

The Hen13, zDEP, FIRE and PREH models are in relatively good agreement
with observational measurements for SFRs larger than $\sim 1
M_{\odot}\,{\rm   yr}^{-1}$, up to $z=1$. At higher redshifts, the
number densities of galaxies with SFRs in the range $1-100 M_\odot/$yr
is over-estimated by up to 0.5~dex. The over-estimation is even larger
for the other models. At all redshifts, the Hen13, zDEP, FIRE models,
and in particular the PREH model predict larger number densities than
the other models for galaxies with SFRs above $>100 M_\odot/$yr, in
better agreement with observations. Only at $z=1.5$, the number
density of these highly star-forming galaxies appears to be strongly
under-estimated with respect to data. 

For galaxies with SFRs lower than $\sim 30 M_\odot/$yr, the Hen13,
zDEP, FIRE and PREH models predict by up to one order of magnitude
smaller number densities than the other models, particularly at high
redshift. This clearly originates from the stronger suppression of
star formation at high redshift in low-mass galaxies (see bottom row
in Fig.  \ref{Baryon_evol}). For galaxies with SFRs lower than ($1
M_\odot/yr$), the Hen13, zDEP, FIRE, and PREH models under-estimate
the measured number densities at all redshifts. We note, however, that
these measurements are based only on SED fitting, and therefore carry
large uncertainties.

\subsection{Specific star formation rates}

\begin{figure*}
  \centering
  \epsfig{file=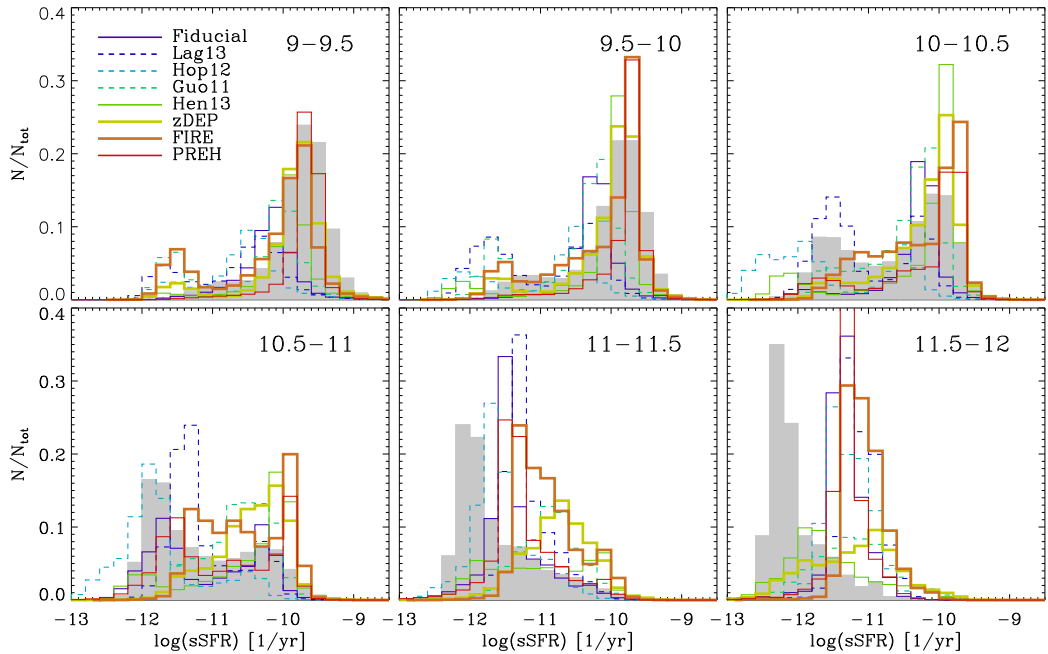,
    width=0.8\textwidth}
  \caption{Distributions of specific SFRs for different stellar mass bins
    (different panels) as predicted by the different stellar feedback models
    considered in this study (colored lines as in
   Fig. \ref{SMF_evol}). Model predictions  are compared with
   observational measurements from SDSS (grey shaded histograms,
   \citealp{Yang07})} 
    {\label{sSFR_distr}}
\end{figure*}

Fig. \ref{sSFR_distr} shows the present-day distributions of the
specific SFR  ($SFR/M_{\mathrm{stellar}}$), as predicted by the
different feedback schemes considered in this study (lines of
different colours). Each panel corresponds to a different bin of
galaxy stellar mass, as indicated by the legend, and model predictions
are compared to observational measurements by the SDSS (grey shaded
histograms).  

For low-mass galaxies ($M_{\mathrm{stellar}} < 10^{10.5} M_\odot$, top
row), the Hen13, zDEP, FIRE and PREH models predict more galaxies with
higher specific SFRs than the other models (whose specific SFRs of
star-forming galaxies are generally too low), in better agreement with
observations. This is clearly a consequence of the delayed star
formation and increasing SFR histories for low-mass galaxies in our
strong ejective (Hen13, zDEP and FIRE) and preventive feedback models
(PREH, see bottom row in Fig. \ref{Baryon_evol}).

For intermediate-mass galaxies $10^{10.5} < M_{\mathrm{stellar}} <
10^{11} M_\odot$ (bottom left panel), the Hen13, zDEP, FIRE and PREH
models tend to predict too large sSFRs, while the others (in
particular the Lag13 and the Hop12 models) provide a better match to
the observational data, with a relatively large fraction of passive
galaxies. 

For more massive galaxies ($M_{\mathrm{stellar}} > 10^{11} M_\odot$),
the specific SFRs are over-estimated in \textit{all} models, although
the disagreement is worse for the Hen13, zDEP, FIRE and PREH
models. We ascribe this failure to the adopted scheme for radio-mode
AGN feedback that, in the framework of our new feedback schemes,
appears to be inefficient in suppressing the star formation in massive
galaxies. A simple parameter change (an increase of the AGN feedback
efficiency) cannot solve this disagreement. We will come back to that
issue in future work.

\subsection{The galaxy main sequence}

\begin{figure}
  \centering
  \epsfig{file=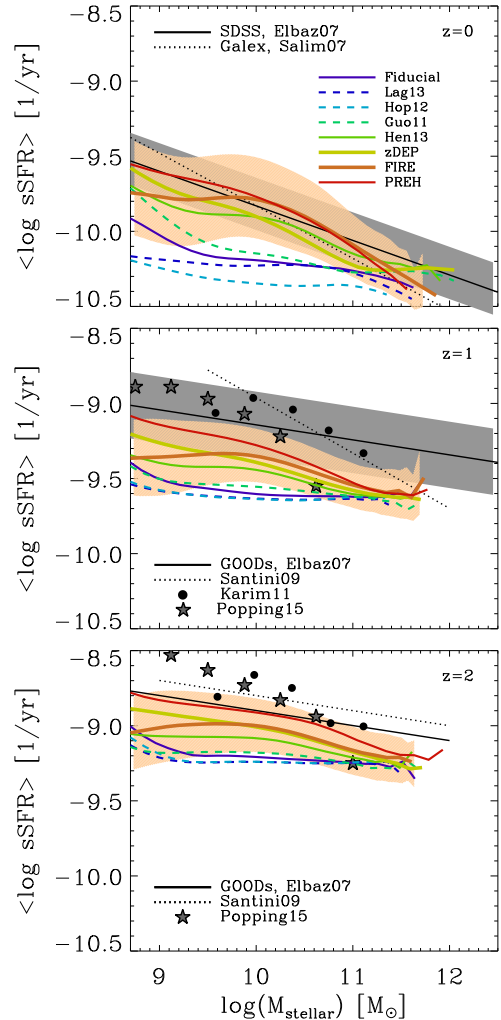,
    width=0.4\textwidth} 
 \caption{Main sequence of star-forming galaxies, i.e.  mean
   specific SFRs are plotted versus the galaxy stellar mass at $z=0,
   1, 2$ (from top to bottom). Model predictions (colored lines and
   shaded areas as in Fig. \ref{Barconv}) are compared to
   observational data (black lines with shaded areas and black symbols -
   \citealp{Elbaz07, Salim07, Santini09, Daddi07}).  }
\label{sSFR-mass}
\end{figure}

Observations reveal a rather tight correlation between (specific) SFRs
of star-forming galaxies and their stellar mass (\citealp{Elbaz07,
  Daddi07, Karim11}), at both low and high redshifts. Some studies
(\citealp{Daddi07, Weinmann12, Granato15}) highlighted that the
state-of-the-art galaxy formation models fail to reproduce this
observational constraint, by under-predicting the normalization of
this correlation and by predicting too shallow slopes for galaxies
with stellar mass lower than $10^{10} M_\odot$ (\citealp{Weinmann12}).

Fig. \ref{sSFR-mass} shows model predictions for the  mean specific SFRs of
  star-forming galaxies (selected as discussed in section \ref{Gas}) as a
function of stellar mass, as predicted by the different feedback schemes used
in this study (lines of different colours; the orange shaded area indicates the
1-$\sigma$-scatter of the FIRE model) at $z=0, 1, 2$ (from top to bottom)
compared to observational data (black symbols, lines and grey shaded area;
\citealp{Elbaz07, Salim07, Santini09, Daddi07}).

In all models, in agreement with the observed trend, the specific SFRs at a
given stellar mass are decreasing with decreasing redshift. The
fiducial, Lag13, Hop12, Guo11 models exhibit the known problem of a
too low  normalisation, and nearly flat or even negative slope of the
relation for low mass galaxies.

In contrast, the Hen13, zDEP, FIRE and PREH models predict larger
specific SFRs at any given mass, and steeper relations. The change in
normalisation with respect to the other models is due to the fact that
there is overall more gas available for star formation. For low mass
galaxies, star formation histories are significantly delayed (see
section \ref{origin}, bottom row of Fig. \ref{Baryon_evol}), leading
to steeper slopes in the these models at redshift zero. The slope of
the relation flattens with increasing redshift, in qualitative
agreement with observational measurements.  

Although model predictions from the Hen13, zDEP, FIRE and PREH models
are definitely in better agreement with data than for the other
feedback schemes tested in this work, they still tend to
under-estimate the specific SFs at high  redshifts. The largest
specific SFRs and the steepest correlations are obtained with the PREH
model. 

\subsection{Quiescent galaxy fractions}

\begin{figure}
  \centering
  \epsfig{file=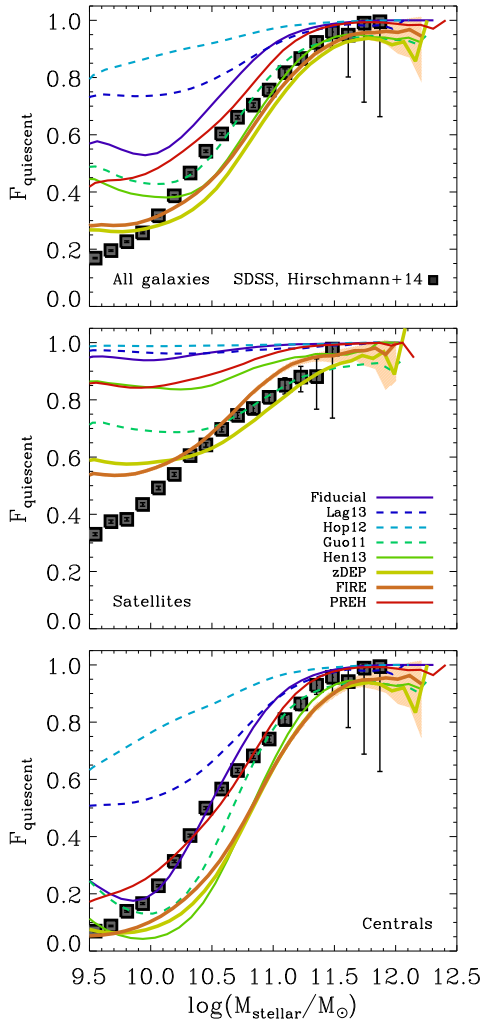,
    width=0.4\textwidth}  
 \caption{ Quiescent galaxy fractions versus galaxy stellar mass for all (top
   panel), satellite (middle panel) and central galaxies (bottom panel). Model
   predictions (colored lines and shaded areas as in
   Fig. \ref{Barconv}) are compared to observational measurements by
   \citet{Hirschmann14a}, based on SDSS (grey symbols). Galaxies
     are selected in both models and simulations assuming the same 
   luminosity cut as in \citet{Hirschmann14a}}
\label{fquiesc}
\end{figure}

One of the long-standing problems of currently used galaxy formation 
models is given by their tendency to largely over-estimate the
fraction of quiescent galaxies, in particular for low-mass satellite
galaxies (see e.g. \citealp{Weinmann09, Weinmann10, Kimm09, DeLucia12,  
  Hirschmann14a}). When first noticed, this model failure was ascribed
to the simplified models for environmental processes: most models
(including the {\sc Gaea} model) assume that the hot gas reservoir
associated with each model galaxy is instantaneously stripped as it is
accreted onto a larger system (i.e. when the galaxy becomes a
satellite). This prevents further cooling and, combined with the
typically efficient stellar feedback, drives a fast suppression of 
the star formation in the model galaxies that transit from star
forming to passive on very short time-scales. While more sophisticated  
treatments of this process  have led to some improvements, model
predictions still appear unable to entirely  reproduce the observed
trends in the local Universe \citep[see e.g.][]{Hirschmann14a,Wang14}.

Fig. \ref{fquiesc} shows the present-day quiescent fractions of all
galaxies (top panel), of satellite (middle panel) and of central
galaxies (bottom panel) compared to observational measurements by
\citet{Hirschmann14a}.  For both   data and models, we have applied
the same luminosity cut of   \citet{Hirschmann14a}. To distinguish
between star-forming and quiescent   galaxies, we again used the
criterion suggested by \citet{Franx08}. We note that the predicted
distributions of sSFR do not agree very well with observational data
(in particular for massive galaxies, which have too high sSFRs). The
adoption of a different separation criterion for quiescent galaxies
would affect the comparison shown in Fig.~\ref{fquiesc}. In
particular, assuming a lower sSFR limit for selecting quiescent
galaxies, would result in a slight under-estimation of the quiescent
fractions for massive galaxies. 

The top panel of Fig. \ref{fquiesc} illustrates that the Hen13, zDEP,
FIRE, PREH and Guo11 models predict quiescent fractions consistent
with observations for galaxies more massive than $10^{10} M_\odot$. In
contrast, the fiducial, Lag13 and Hop12 models over-estimate the
fraction of quiescent galaxies, particularly at low masses. For
galaxies less massive than $10^{10} M_\odot$, the quiescent galaxy
fractions are over-estimated in all models to a very different degree. 

In the middle and bottom panels, we split the galaxy populations into
centrals and satellites. Regarding centrals, the fiducial, PREH and
Guo11 models predict realistic quiescent fractions, while the strong 
ejective models (Hen13, zDEP and FIRE) predict slightly too few
quiescent galaxies -- an effect already discussed in
Fig. \ref{sSFR_distr} and likely pointing towards more fundamental
modifications of the modelling adopted for radio-mode AGN feedback. 

For satellites, the quiescent fractions are found to be very strongly
dependent on the stellar feedback scheme. We find reduced quiescent
fractions in the strong ejective schemes (Hen13, zDEP, FIRE) and in
the Guo11 models with respect to the other models. We stress that, in
all adopted schemes, we assume an instantaneous stripping of the hot
gas reservoir associated with infalling galaxies. Therefore, the model
quiescent satellite fractions are mainly regulated by: (i) the amount
of cold gas at the time of infall, and (ii) the rate at which the cold
gas gets re-heated and is, thus, blown out of the satellites (later
on, it can be re-incorporated only onto the central galaxy of the halo).

At the time of infall, the (central) galaxies  in the strong ejective
feedback models (Hen13, zDEP, FIRE) are more star-forming, i.e. they
have a larger cold gas content, which allows them to sustain star
formation for longer time scales. The quiescent fractions predicted by
the strong ejective feedback schemes vary significantly: the more gas
gets re-heated (as e.g. in the Hen13 model, see top row in
Fig. \ref{Baryon_evol}), the less gas is available for further star
formation, the faster is the remaining cold gas consumed, and the
higher are the quiescent fractions. This explains the larger quiescent
fractions in the Hen13 model (characterized by almost constant
re-heating rates) with respect to the FIRE and zDEP models
(characterised by decreasing re-heating rates towards $z=0$). The
rather low quiescent fractions in the Guo11 model can be explained
with similar arguments: the re-heating rates in this model are almost
constant and rather low with respect to the Hen13 model or even the
fiducial model.  

We stress that {\it all} our models are based on the simplified
assumption of an instantaneous hot gas stripping of satellites. 
Therefore, in the framework of our models, the fractions of
  passive galaxies (both centrals and satellites) are primarily
  determined by internal physical processes, with environmental
  processes playing only a secondary role. The fact that the most
successful models are still over-estimating the lowest mass quiescent
satellites suggests that environmental effects become important at
these mass scales. We postpone a more detailed analysis  to a future
work (Hirschmann et al., in prep.).

%******************************************************************************
%******************************************************************************
\section{Stellar populations}\label{Stellarpop}
%******************************************************************************
%******************************************************************************

In this section, we discuss the impact of the different stellar
feedback schemes adopted in this study on the stellar populations of
our model galaxies, particularly with respect to their stellar
metallicities, ages and colours. 

\subsection{Stellar metallicity}

\begin{figure}
  \centering
  \epsfig{file=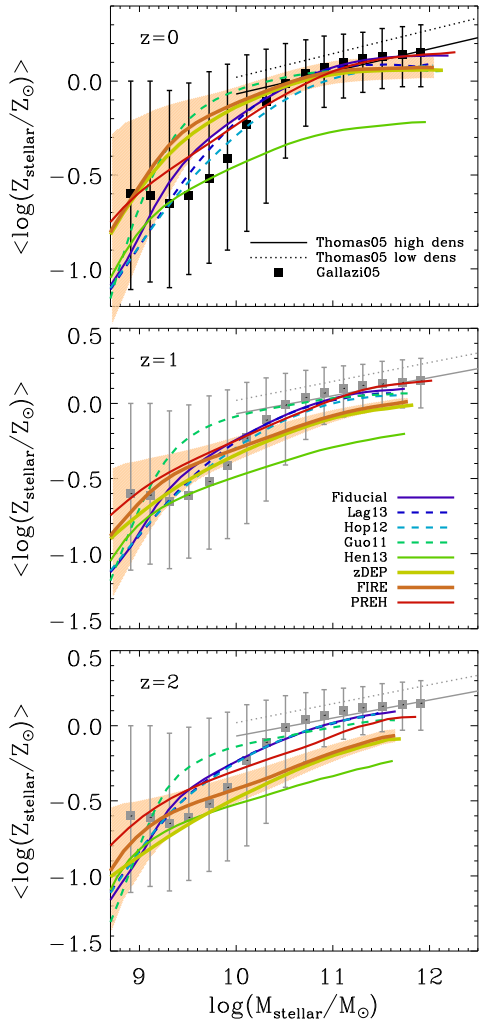,
    width=0.4\textwidth}
 \caption{ Mean stellar metallicity versus galaxy stellar
   mass at $z=0, 1, 2$ (from top to bottom). Model predictions
   (colored lines and shaded areas as in Fig. \ref{Barconv}) are
   compared to observational data of the present-day Universe
   (black/grey lines with shaded areas and black/grey symbols,
 \citealp{Thomas05, Gallazzi05}).  }  
\label{Stellarmet}
\end{figure}

Fig. \ref{Stellarmet} shows the stellar mass-stellar metallicity
relation at $z=0, 1, 2$ for our different feedback schemes, compared
to the present-day  observed relation (black symbols and lines,
corresponding to measurements by \citealp{Gallazzi05} and
\citealp{Thomas05} respectively). For reference, we show the $z=0$
observational measurements also in the panels corresponding to higher
redshifts. At $z=2$, the fiducial, Lag12, Hop11, Guo11 models predict
that galaxies more massive than $10^{10} M_\odot$ have already reached
super-solar metallicities, and their metal content is hardly evolving
down to $z=0$. The PREH model shows a similar behaviour: since it
includes the same  re-heating and ejection scheme as in the fiducial
model, we expect a similar metal enrichment history.

Strong ejective feedback (Hen13, zDEP, and FIRE illustrated by green,
yellow and orange solid lines) can have a significant impact on the
evolution of the stellar metallicity: early star formation is
suppressed and, in addition, metal-enriched gas is ejected out of the
galaxy. As a consequence, metal enrichment of the stellar component is
delayed, so that all galaxies in the strong ejective feedback models
have sub-solar metallicity at $z=2$. In the FIRE and zDEP models, the
stellar metallicity is strongly increasing with decreasing redshift,
approaching super-solar metallicity for the most massive galaxies, in
good agreement with the present-day observed relation. In contrast,
the Hen13 model predicts a too strongly delayed metal enrichment so
that even the most massive galaxies are always below solar levels.  

This different behaviour of the strong ejective feedback schemes
(Hen13, zDEP, and FIRE) can be explained considering the larger
re-heating and ejection rates in the Hen13 model with respect to the
FIRE and zDEP models below $z\sim2$ (see two first rows in
Fig. \ref{Baryon_evol}). Since the metal flows are assumed to follow
the gas flows, more metals are ejected from the cold, star-forming gas
phase and transferred to the hot gas phase at late times in the Hen13
model than in the FIRE and zDEP models. Due to mixing with less metal
enriched gas in the hot gas phase (or in the ejected phase), the
metallicity of the cooled gas is typically lower than that of the cold
gas. This tends to dilute the metallicity of the cold gas component,
and therefore keeps the metallicity of the stars formed later low.

\subsection{Galaxy stellar ages}

\begin{figure}
  \centering
  \epsfig{file=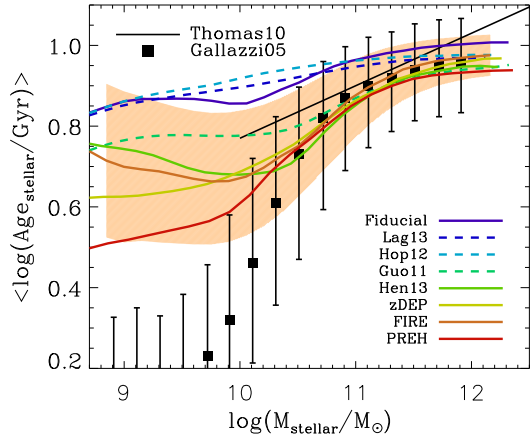, width=0.45\textwidth}
  \caption{Present-day mean stellar ages as a function of the
    galaxy stellar mass in the different feedback models considered in
    this study (coloured lines and shaded areas as in
    Fig. \ref{Barconv}), compared to observations by \citet{Thomas10}
    and \citet{Gallazzi05} (black solid line and symbols).}
  {\label{Stellarages}} 
\end{figure}

Fig. \ref{Stellarages} shows the present-day stellar mass-age relation for the
different feedback models (lines of different colours; the orange shaded area
indicates the 1-$\sigma$-scatter for the FIRE model), compared to observational
measurements by \citet{Thomas10} and \citet{Gallazzi05} (black solid line and
symbols). Both model stellar ages and observational measurements are
  luminosity weighted (r-band, SDSS).

Galaxies in the fiducial, Lag13 and Hop12 models are generally too old
compared to observational measurements. In the Guo11 feedback scheme, 
galaxies are on  average younger, and in good agreement with
observations for galaxies more massive than $10^{10.5} M_\odot$. Less
massive galaxies are still too old compared to data. The younger
stellar ages in the Guo11 model are a natural consequence of the
strongly suppressed star formation at early cosmic times, compared to
the other models. 

The trend is even more extreme for the Hen13, zDEP, FIRE and PREH
models, resulting in fairly realistic stellar ages for galaxies with
masses above  $10^{10} M_\odot$. For less mass galaxies ($ < 10^{10}
M_\odot$), stellar populations tend to be still too old compared to
measurements by \citet{Gallazzi05}, although the scatter in both
observations and models at the low mass is rather large (+/-
0.2~dex). 

The PREH and zDEP models predict the youngest low mass galaxies with 
respect to the other feedback models. This is due to the stronger
early star formation suppression (see bottom panel in
Fig. \ref{Baryon_evol}) with respect to the other successful FIRE or
Hen13 models. Interestingly, for the FIRE and Hen13 low mass galaxies,
stellar ages are increasing with decreasing stellar mass, a trend that
is neither visible in the zDEP and PREH models and, nor in the
observations.

\begin{figure*}
  \centering
  \epsfig{file=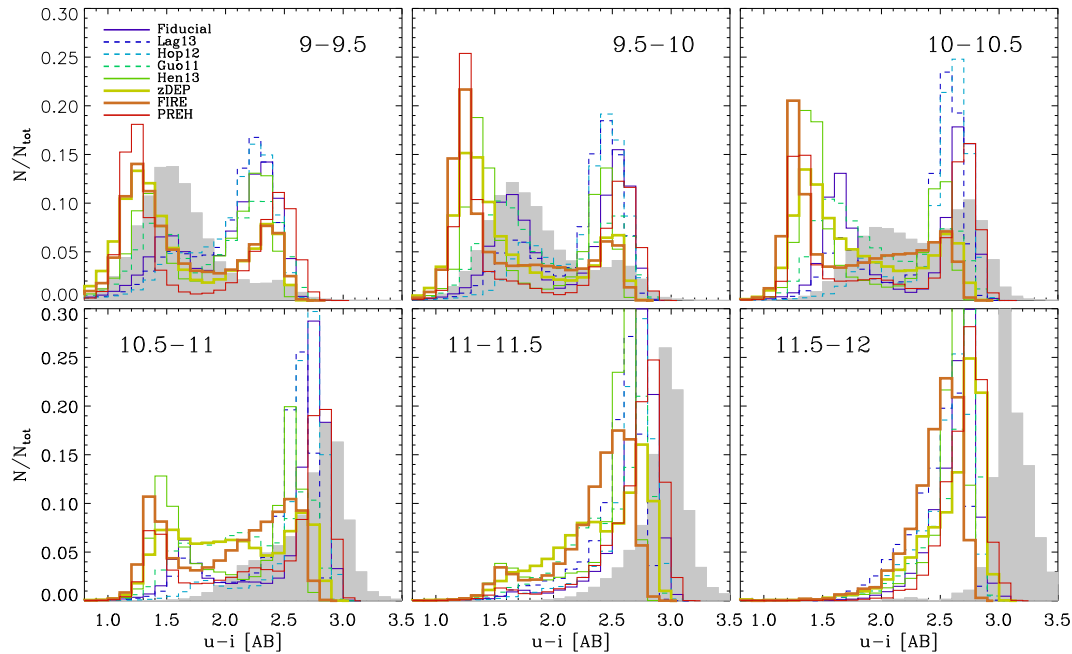, width=0.8\textwidth}
  \caption{Present-day u-i color distributions for different stellar mass bins
    (different panels) as predicted by the different stellar feedback
    models (colored lines as in Fig. \ref{SMF_evol}),
    compared to SDSS measurements (grey shaded histograms,
    \citealp{Yang07}).}  {\label{Colordistr}} 
\end{figure*}

\subsection{Galaxy colours}

We conclude our sections on model results by showing the predicted
colour distributions from the different feedback schemes used in our
study. While these are direct observables, assumptions need to be made
in the models to predict them from physical properties (e.g. on
stellar population models, dust attenuation, etc.). Model predictions
for the u-i colour distributions are shown in Fig. \ref{Colordistr},
and compared with measurements from the SDSS  (grey shaded
histograms). 

All models predict a rather strong colour bi-modality for galaxies less
massive than $10^{11} M_\odot$, while in the data this is evident only
for galaxies in the mass range $10^{9.5}-10^{10.5} M_\odot$. The
colours of blue model galaxies tend to be generally too blue (see top
row). For galaxies with stellar mass $<10^{11} M_\odot$, the Hen13,
zDEP, FIRE, and PREH models predict a larger fraction of blue galaxies
than the other models, consistently with the lower stellar ages
discussed above. This results in a better agreement with observational
measurements for masses $<10^{10} M_\odot$, in particular when
considering the FIRE and zDEP models. For galaxies with masses between
$10^{10}-10^{11} M_\odot$, the fraction of blue galaxies is, however, clearly
over-estimated with respect to data. 

For galaxies with mass larger than $10^{11} M_\odot$, model colours are
generally not red enough compared to observational measurements, a
trend that appears even worse for the Hen13, zDEP, FIRE and PREH
models. This is consistent with the non negligible levels of star
formation in massive galaxies discussed earlier (see
e.g. Fig. \ref{Baryon_evol} bottom, left panel). These are low
enough to not affect significantly stellar masses, ages and
metallicities, but have a larger effect on colours (especially in the
blue bands).

%**********************************************************************
\section{Discussion}\label{discussion}
%**********************************************************************

In the last decade, a number of theoretical studies have pointed out the
existence of a {\it fundamental problem} with the evolution of low mass ($\leq
10^9\, {\rm M}_{\odot}$) galaxies in hierarchical models. This problem has
different manifestations: (i) models tend to systematically and significantly
over-predict the number density of low-mass galaxies  (e.g. \citealp{Guo11,
    Hirschmann12b, Lu14}); (ii) low mass galaxies, and in particular satellite
galaxies, tend to be too passive with respect to observational data 
  (e.g. \citealp{Kimm09, Weinmann10}). The work carried out in the last decade
has pointed out that these problem cannot be overcome by simple modifications
of the satellite treatment like a non instantaneous stripping of the hot gas
reservoir associated with infalling galaxies (\citealp[see
    e.g.][]{Hirschmann14a}), or a more efficient stellar feedback or a stronger
scaling with halo circular velocity ( \citealp[e.g.][]{Guo11}).

\subsection{Comparison with previous work}\label{SAMcomp}

A few solutions to the problems mentioned above have been proposed in 
the framework of different theoretical models of galaxy formation and
evolution. We discuss them below, comparing results from these studies 
to ours. 

\subsubsection{{\sc L-Galaxies} models} 

Although different in a number of details, our {\sc Gaea} model
originates from the same model as the most recent versions of that
built by the  `Munich' group: {\sc L-Galaxies}. 

The over-estimation of the low-mass galaxies at $z>0$   in the Guo11
feedback scheme adopted in {\sc Gaea} is entirely consistent with
results presented in \citet{Guo11}, despite significant differences in
modelling the metal enrichment and the evolution of satellite
galaxies. This scheme also leads to low mass galaxies that are too
red, and massive galaxies that are  too blue compared with
observational data.   

Although both our Guo11 feedback scheme and the original 
\citet{Guo11} model predict reduced, and thus more realistic,
present-day quiescent satellite fractions at fixed stellar mass, the  
physical reasons for this are different: the lower quiescent fractions
in the Guo11 model in {\sc Gaea} with respect to  our fiducial model 
are mainly due to the lower re-heating efficiency, which we found to
be necessary for predicting a realistic metal content in our
present-day galaxies. In contrast, we suspect that in \citet{Guo11},
the reduced quiescent satellite fractions are mostly due to the
different treatment of environmental effects (a delayed stripping of
the hot gas associated with infalling satellites). In fact, when
assuming the same re-heating efficiency as in \citet{Guo11}, we obtain
much larger quiescent fractions with {\sc Gaea}. 

In the most recent version of the {\sc L-Galaxies} model,
\citet{Henriques13} and \citet{Henriques15}, have proposed a modified 
gas reincorporation scheme where the re-incorporation time-scales
vary with the inverse of the halo mass. This new element
introduced in the feedback scheme is able to bring model predictions
in agreement with the measured evolution of the GSMF. Our
implementation of the Hen13 feedback scheme in the {\sc Gaea} model is 
fully consistent with results discussed by \citet{Henriques13}. The
physical properties of low-mass galaxies are in better agreement
with data: they are younger, bluer, more star-forming, more gas-rich
and the fraction of quiescent satellites is reduced. Nevertheless,
some inconsistencies with observations remain, e.g. with respect to
colour and sSFR distributions, in particular for the most massive
galaxies that tend to be too blue and star forming.

We find, however, significant differences between our {\sc Gaea}
implementation of the Hen13 feedback scheme and results by
\citet{Henriques15} in the framework of the {\sc L-Galaxies} code. In
the former, the present-day stellar (and gaseous) metallicity is
severely under-estimated compared to observational measurements, while 
\citet{Henriques15} show a very good agreement with data. Their good
match is due to an unrealistically high metal yield (0.047) adopted in
the framework of their instantaneous recycling approximation. In the
{\sc Gaea} model, the metal yield is no longer a free parameter
anymore, but is determined by the set of metal yields chosen and,
although there are some uncertainties in the yields of some elements,
it cannot be significantly larger than $\sim 0.02$. In this way,
observational measurements of the metal content in the local Universe
represent a truly independent constraint with respect to measurements
of the GSMF. 

We find that, reducing the re-heating and ejection efficiencies in our
Hen13 feedback implementation such that massive galaxies reach more
realistic super-solar metallicities, the evolution of the low-mass end
of the SMF cannot be reproduced anymore, and the number densities of
low-mass galaxies are again over-estimated with respect to data. 
  Our analysis does not exclude that it   is possible to find a good
  match with the observed evolution of the GSMF and   the measured
  metal content of galaxies in the local Universe with a lower
  chemical yield, within the \citet{Henriques15} scheme. Such an
  analysis is   beyond the aims of this paper, but can be effectively
  carried out using the   Monte Carlo Markov Chain approach used in
  Henriques et al. If such a solution  is found, it will be
  interesting to compare predictions from such a model with results
  shown in our Fig. 4. 

%% Our results suggest that a decreasing mass loading of the
%% reheated and ejected gas over cosmic time represent a necessary
%% condition to match simultaneously the observed evolution of the GSMF
%% and the observed metal content of galaxies in the local Universe. 

\subsubsection{Independent semi-analytic models}

\citet{Lagos13} argued that with their feedback model they can reproduce the
evolution of the shallow end of the luminosity function (they do not show a
comparison between model predictions and the observed GSMF).  For
  our Lag13 implementation in {\sc Gaea}, the faint end of the
  luminosity function is also shallower than in our fiducial
  model. Model results are, however, inconsistent with the observed
  evolution of the GSMF. However, it should be noticed that the model
  presented in \citet{Lagos13} does not follow explicitly the fate of
  the re-heated gas. One difference between our {\sc Gaea}
implementation and the original \citet{Lagos13} model is that the
latter includes an explicit modelling of the HI and H$_2$
formation. This, however, is not expected to introduce significant
differences in model predictions \citep{Lagos11}. Therefore, our
results are probably due to the overall significant differences
between the modelling of physical processes in {\sc   Galform} and
{\sc Gaea}, highlighting the strong interplay between stellar feedback
and other physical processes driving galaxy evolution.

Our simple implementation of pre-heating agrees with results based on
the more sophisticated model discussed in \citet{Lu15}: reducing the
amount of newly infalling gas can lead to realistic SFR histories,
cold gas fractions and baryon conversion efficiencies in low mass
galaxies. \citet{Lu15} do not discuss the impact on metallicity, that
we find to be a weak point of our pre-heating implementation: this
model predicts a too fast enrichment of the cold gas content, in
contrast with observational measurements. We might speculate
that, in reality, preventive and ejective feedbacks are both at play,
likely with different relative importance at different cosmic epochs
and at different mass scales.

\citet{White15} study the impact of preventive and redshift dependent
ejective feedback models using the semi-analytic model presented in
\citet{Somerville12}. They base their work on `ad-hoc'
parametrizations (just like our zDEP model) aimed at matching the
number density of low-mass galaxies. Even if their specific feedback
implementations differ from ours in detail, they also find that both
ejective and preventive feedbacks reduce the number density of low-mass
galaxies and predict larger cold gas fractions and specific SFRs, in
fairly good agreement with observational constraints.  In their
preventive model, the gaseous metallicity at a fixed stellar mass is
fairly constant over time (like in our implementation), and only the
ejective feedback predicts a (slightly) increasing gaseous metal
content. Although  not in perfect agreement with observational
measurements, their results and conclusions are consistent with
ours.

 \citet{Mitchell14} show that a redshift dependent reincorporation 
  time-scale provides a viable solution to obtain more realistic
  stellar mass   assembly histories and a negative slope in the
  sSFR-stellar mass relation   (particularly for low mass
  galaxies). Their assumption, however, results in a   worse match
  with the GSMF than a simple dependence on halo mass as suggested by
  \citet{Henriques13}. We can speculate that, within our \textsc{Gaea}
  model, a redshift dependent reincorporation time-scale (instead of a
  redshift  dependent outflow rate) can, in some form, provide a good
  match to the observed evolution of the GSMF. In this respect, it
  would be beneficial to study how the reaccretion time-scale vary ad
  a function of redshift in the state-of-the-art cosmological
  hydrodynamical simulations.

\subsubsection{Cosmological hydrodynamical simulations}

Over the last years, several studies, based on different
hydrodynamical simulations codes, have been successful in reproducing
the estimated baryon conversion efficiencies and, in a few cases, even
in matching the shallow low-mass end of the GSMF (\citealp{ Stinson13,
  Aumer13,   Furlong15, Hopkins14}). This success was, however,
achieved by using fundamentally different sub-resolution models and by
making different assumption for stellar feedback. This is just a
reflection of the intrinsic difficulty to model stellar feedback,
i.e. processes such as stellar winds, radiation pressure, ionising
radiation, expansion of SN bubbles,  from `first principles'. 

In agreement with results discussed above, all successful simulations
predict that, at high redshift, the vast majority of gas should be
prevented from cooling and forming stars, either by ejecting accreted,
low angular momentum gas or by preventing gas infall from the IGM.
The consequences of having large amounts of non-cooling gas are
similar to those discussed earlier: star formation is suppressed at
high redshift and delayed towards later times, metal enrichment is delayed,
cold gas fractions are increased. In particular, low mass-galaxies are
bluer, more star-forming (higher sSFR) and younger
(e.g. \citealp{Aumer13,    Hirschmann13, Hopkins14}), in overall
better agreement with observations. 

Some simulations (e.g. \citealp{Hopkins14}), however, show that when
assuming some form of strong ejective feedback to reproduce the
  observationally   inferred mass assembly, they fail to generate
simultaneously realistically looking, thin disk-like galaxies. This
tension may point towards other, less `violent' mechanisms to get rid
of large amounts of cold gas in galaxies or to prevent pristine gas
from infall in order to suppress early star formation by {\it
  simultaneously} allowing for a build-up of a realistic
spiral/disk-like structure.

\subsection{A successful feedback scheme in {\sc Gaea}. Limits and
  ways forward.} 

In our analysis, four of the feedback schemes tested, the Hen13, zDEP,
FIRE and PREH models, are reasonably successful in reproducing the
observed trends in galaxy mass assembly. This is achieved by `storing'
large amounts of gas in an `ejected gas reservoir', where it is not
available for cooling. These models predict specific SFRs, stellar
ages and cold gas fractions that are overall in better agreement with
observational measurements than our fiducial feedback scheme used in
previous work. When considering other observational constraints, such
as the gaseous and stellar metal content and the quiescent satellite
fractions, our zDEP and FIRE models appear to perform better. In
particular, both the Hen13 and PREH scheme predict too many passive
satellite galaxies, and the Hen13 model predicts a too efficient
suppression of the metal enrichment so that the metal content of model
galaxies is under-estimated, particularly in the local Universe. The
PREH model can match the observed present-day mass-metallicity
relation, but it results in a too fast enrichment of the cold gas
content in massive galaxies. We stress that, given the self-consistent
treatment of chemical enrichment in the {\sc Gaea} model, the chemical
yield cannot be regarded as a free parameter and observational
measurements of the metal content of different baryonic components
represent independent and powerful constraints to distinguish among
different feedback schemes.

Both favoured feedback models (FIRE and zDEP) imply an explicit
redshift dependence of the mass-loading for the re-heated and ejected
gas, at fixed circular velocity. In particular, they predict that a
baryon fraction of $\sim$ 60-70 per cent is not available to cool onto
the central galaxy disk and to form stars at high
redshift. Forthcoming measurements of the gas composition `outside'
galaxies, e.g. in the CGM and/or IGM, will help to falsify model
predictions. Specifically, the metal content of the gas in the
surroundings of a galaxy will provide strong constraints on the
feedback schemes: while purely preventive models would predict a
rather pristine gas composition, ejective models favour a
metal-enriched CGM. 

In spite of the success of our FIRE and zDEP models, some
inconsistencies with observations remain. In particular, massive
galaxies ($>10^{11} M_\odot$) have still too high SFRs at low
redshifts. As mentioned, this is most likely related to the currently
adopted model for AGN feedback. We find, however, that simply
increasing the corresponding feedback efficiency is not sufficient to
bring model predictions in better agreement with data. This points
towards a more fundamental revision of the AGN feedback scheme with
the inclusion of e.g. the effect of quasar driven winds
(e.g. \citealp{Ostriker10}), that is currently neglected in the {\sc
  Gaea} model. 

In addition, in the FIRE model, low-mass galaxies ($<10^{9.5}
M_\odot$) still tend to be too numerous at $z=0.5-2$, too red and too 
old. We argue that, at this mass scale, environmental effects become
important and a better treatment of satellite galaxies (e.g. a non
instantaneous stripping of the hot gas reservoir associated with
infalling satellites) might improve the agreement with observational
data. We plan to address this in future work. It is interesting to
note that these problems are alleviated in the zDEP model and in the
PREH model. It remains to be seen if a more sophisticated preheating
model, coupled with a strong ejective feedback scenario, could provide
a better description of the low-mass galaxy populations. 

Our analysis does not allow us to further discriminate between the
FIRE and zDEP model. As the latter has been constructed `ad hoc' to
reproduce the evolution of the GSMF, we regard the former as our
`reference model'. Although this model is based on results from
cosmological hydrodynamical simulations that include an explicit
modelling of complex and relevant physical processes, we stress that
this is still a `sub-grid' (semi-analytic) model.  

%**********************************************************************
\section{Summary}\label{summary}
%**********************************************************************

In this study, we carried out a systematic analysis of the influence
of different stellar feedback models, with a focus on the
mechanisms/schemes required to reproduce the observed evolution of the   
GSMF. Our work is based on the {\sc Gaea} model, an evolution of the
semi-analytic model presented in \citet{DeLucia07}, and includes the
detailed chemical enrichment scheme  introduced in \citet{DeLucia14}.
For each feedback model, we have adjusted the corresponding parameters
(for stellar feedback and the reincorporation efficiency, see table 1)
so as to match the exponential cut-off of the stellar mass function at
z = 0, and by simultaneously trying to obtain a good match with the
observed mass-metallicity relation in the local Universe and the
measured evolution of the galaxy stellar mass function (GSMF) at
higher redshift. 

Our main results can be summarised as follows.
\\

(i) Both a strong ejective (Hen13, zDEP, FIRE) and a preventive
  form of feedback (PREH) are capable of reproducing the observed
evolution of the GSMF. In our successful strong ejective feedback
  models (zDEP, FIRE), the mass-loading is dependent on redshift, and
large fractions of the hot gas associated with the parent dark matter
halo can be driven out of the halo (ejected) and made unavailable for
cooling for relatively long time-scales. In order to reproduce
  the observed trends, large amounts of baryons (up to 70 per cent of
the baryon budged available) have to be unavailable for cooling,
particularly at high redshift. Delayed gas re-incorporation later
reduces the large amounts of cold, star-forming gas at high redshifts
(as already suggested by \citealp{Henriques13}). The same
  schemes predict bluer colours and larger amounts of cold gas for
  low-mass   galaxies, in overall better agreement with observational
  measurements than   our previous fiducial model. \\

(ii) Due the full chemical enrichment scheme in our {\sc Gaea} model,
we can use observational measurements of the gaseous and stellar
metallicity content of galaxies as independent constraints. A feedback
model with strong but constant gas outflows and a delayed
gas-recycling (as suggested by \citealp{Henriques13}) leads to a too
efficient suppression of the metal enrichment, in contrast with
observational data. Our preventive feedback   scheme (PREH) is
able to predict a realistic present-day metal content, but the ISM is
enriched too quickly.  Only by using a strong ejective feedback
  with a redshift dependent mass-loading (zDEP, FIRE), can we
successfully reproduce both a delayed metal enrichment and a realistic
present-day stellar and gaseous metallicity.\\ 

(iii) Although significantly improved with respect to previous
results, our new models are not without problems. In particular,
massive galaxies  appear to be too active with respect to
observational measurements at low redshift and the number densities of 
low-mass galaxies are still over-estimated over the redshift range $z
= 0.5 - 2$. Dwarf galaxies also appear too old over this redshift
range. We argue that these problems require a significant revision of
our AGN feedback model, and a more sophisticated treatment of
environmental processes. We plan to address these issues in future
work. \\  

Despite significant recent progress in reproducing a number of crucial
observed trends, current galaxy formation models are still highly
degenerate in terms of the adopted stellar and recycling schemes. More
and more stringent observational constraints are coming in the next
future, for example through precise high redshift measurements of the
stellar mass functions, baryon conversion efficiencies and stellar
metallicities. Strong constraints on the baryon cycle can come from
observations of the diffuse gas and metals in the GCM and IGM.

\section*{Acknowledgements}

Galaxy catalogues for our new {\sc Gaea} model (with the FIRE feedback
scheme) implemented on the Millennium merger trees will be publicly
available at http://www.mpa-garching.mpg.de/millennium.

We thank Rachel Somerville, Pierluigi Monaco, St\'ephane Charlot and
Simon White, Bruno Henriques and Rob Yates for fruitful discussions.
We thank Volker Springel and Mike Boylan-Kolchin for making available
to us the merger trees from the Millennium and Millennium II
simulations and Gerard Lemson and Simon White for helping to make our
model catalogues publicly available.

MH acknowledges financial support from the European Research Council
via an Advanced Grant under grant agreement no. 321323 NEOGAL.  GDL
and MH acknowledge  financial support from the European Research
Council under the European Community's Seventh Framework Programme
(FP7/2007-2013)/ERC grant agreement n. 202781. GDL acknowledges
support from the MERAC foundation.  FF acknowledges financial support
from the grants PRIN INAF 2010 `From the dawn of galaxy formation',
and PRIN MIUR 2012 `The Intergalactic Medium as a probe of the growth
of cosmic structures'.

\bibliographystyle{mn2e}
\bibliography{Literaturdatenbank}

\label{lastpage}

\begin{appendix}

%%%%%%%%%%%%%%%%%%%%%%%%%%%%%%%%%%%%%%%%%%%%%%%%%%%%%%%%%%%%%%%%%%%%%%%%%%%%%
\section{Milky-Way like galaxies}\label{MWappendix}

%%%%%%%%%%%%%%%%% Aquarius criteria %%%%%%%%%%%%%%%%%%%%%%%%%%%%%%%%%%%%%%%%

\begin{figure}
 \centering \vspace{1.0cm}
 \epsfig{file=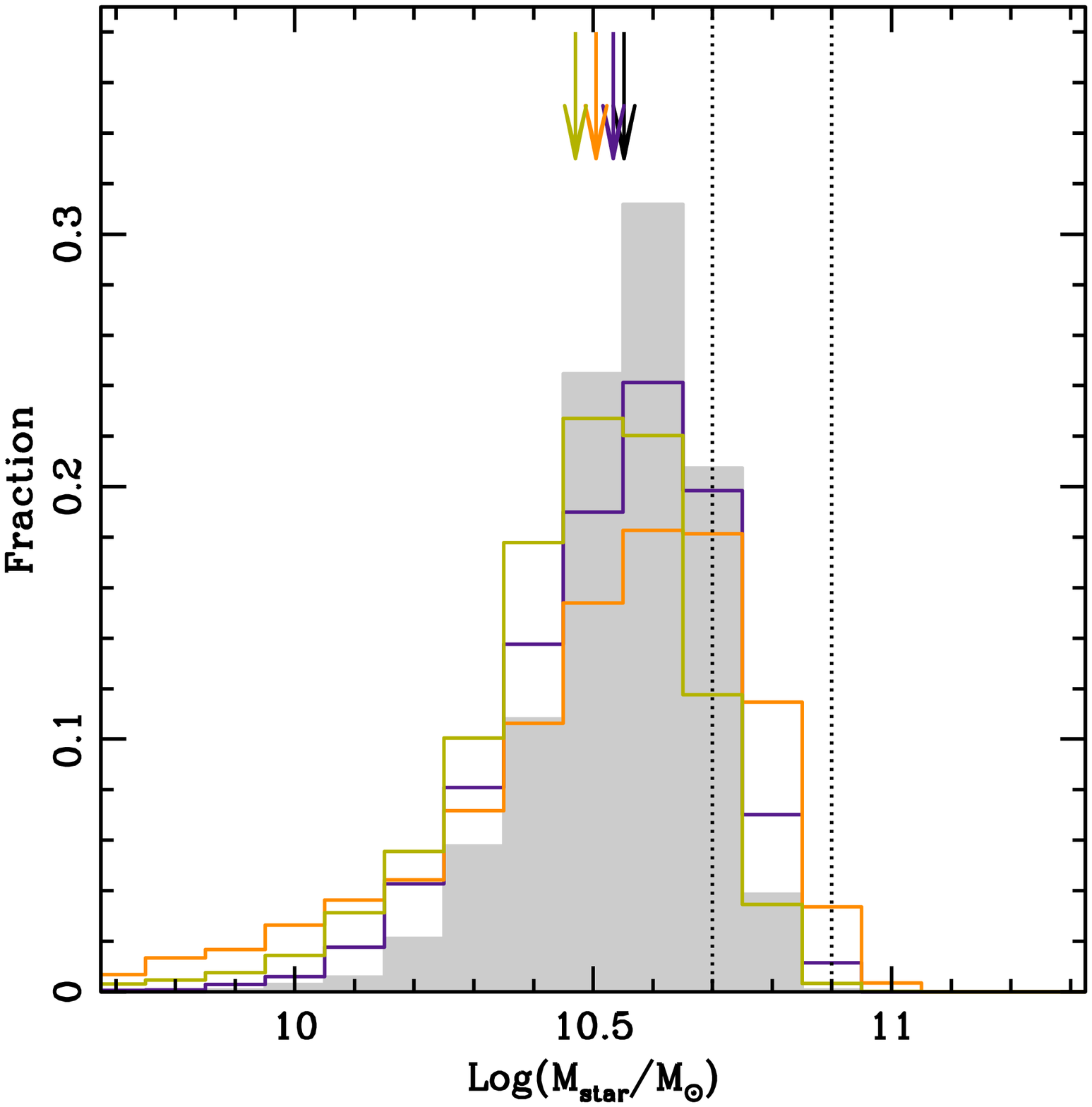,width=0.33\textwidth}\vspace{1.5cm}
% {\label{sfr_MW}} 
%\end{figure}
%\begin{figure}
% \centering
 \epsfig{file=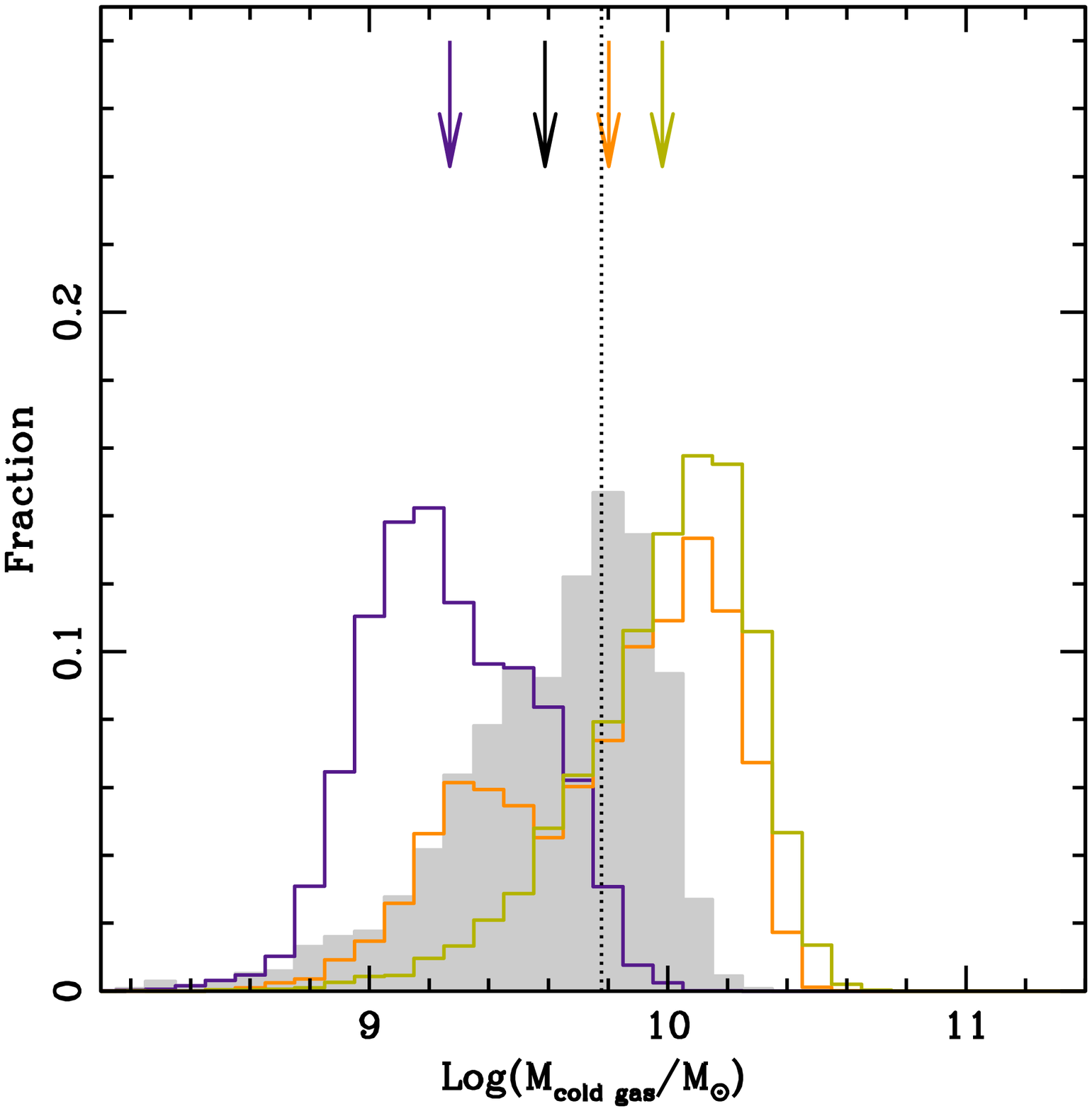,width=0.33\textwidth}\vspace{1.5cm}
 \epsfig{file=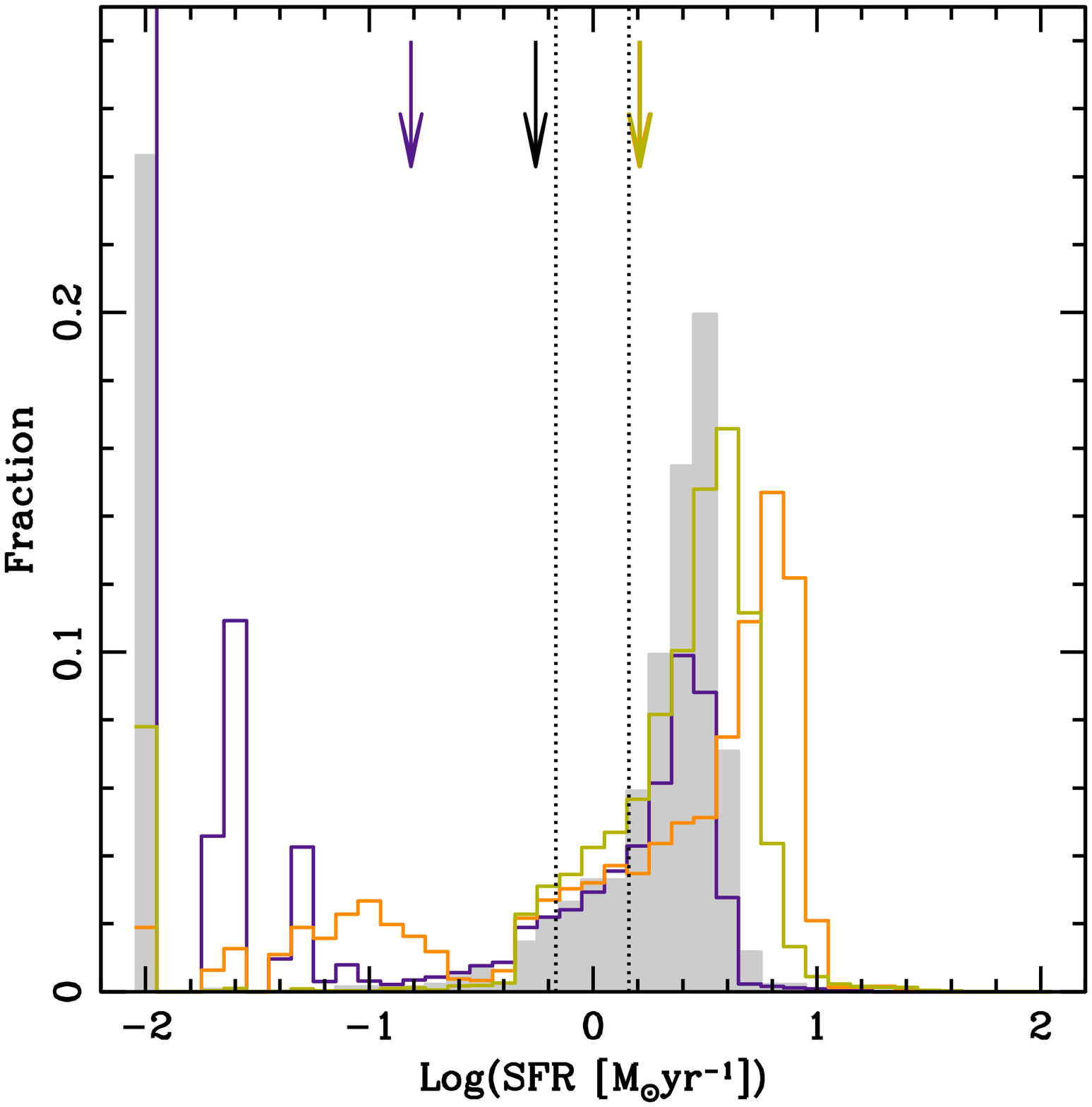,width=0.33\textwidth} 
 \caption{Distributions of galaxy stellar mass (top panel), cold gas mass
   (middle panel) and SFRs (bottom panel) for Milky-Way like galaxies selected
   following the Aquarius-like criteria described in the text. The grey
   shaded area shows results obtained using the model described in
   \citet{DeLucia14}, while the coloured histograms correspond to the
   fiducial (lila), FIRE (orange), and zDEP (yellow) models discussed in this
   work. Arrows correspond to the median of the distributions, while the
   vertical dotted lines show the (range of) observational estimate(s).}
         {\label{sfr_MW}} 
\end{figure}

\begin{figure}
 \centering \vspace{1.0cm}
 \epsfig{file=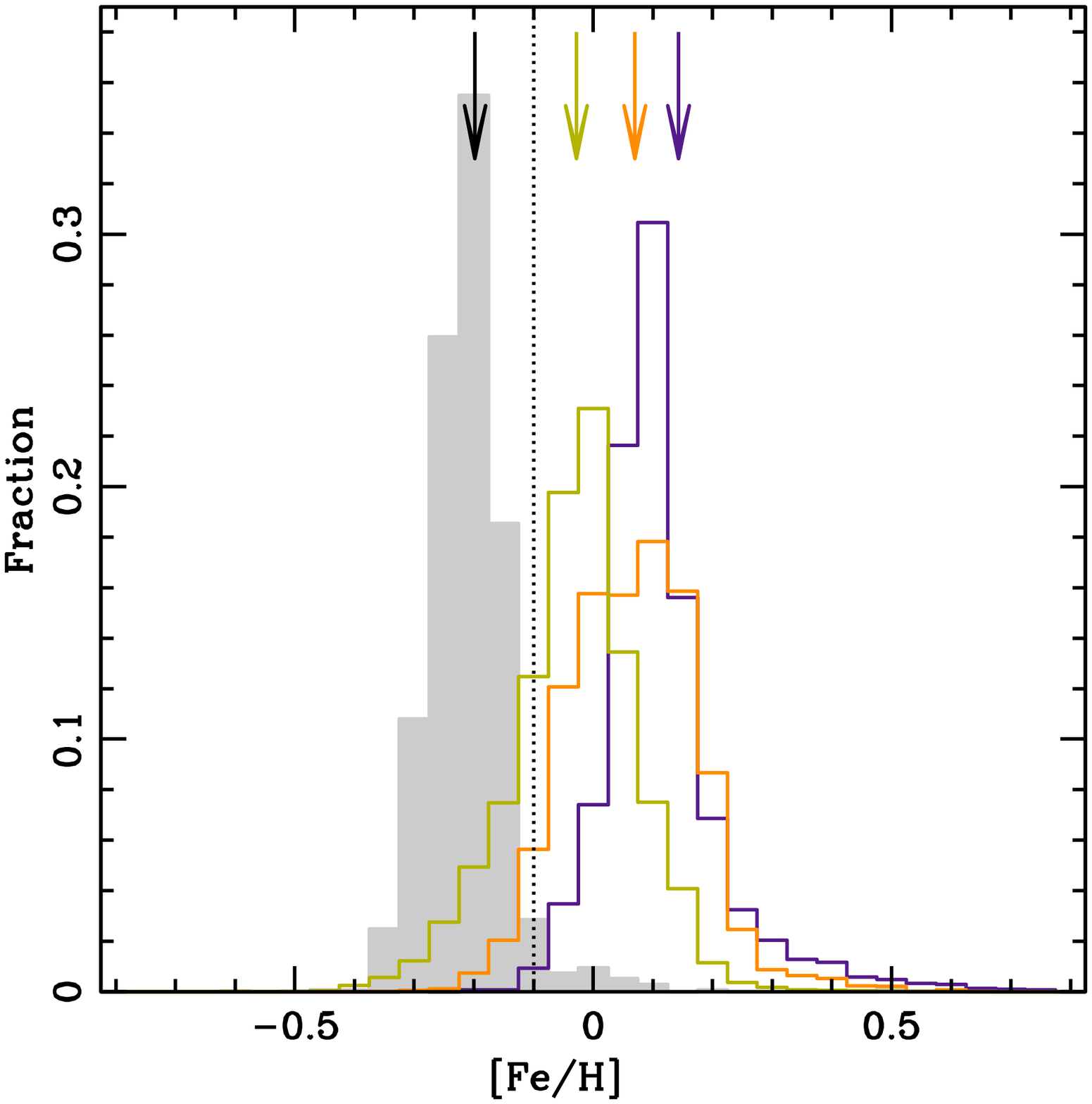,width=0.33\textwidth}\vspace{1.5cm}
 \epsfig{file=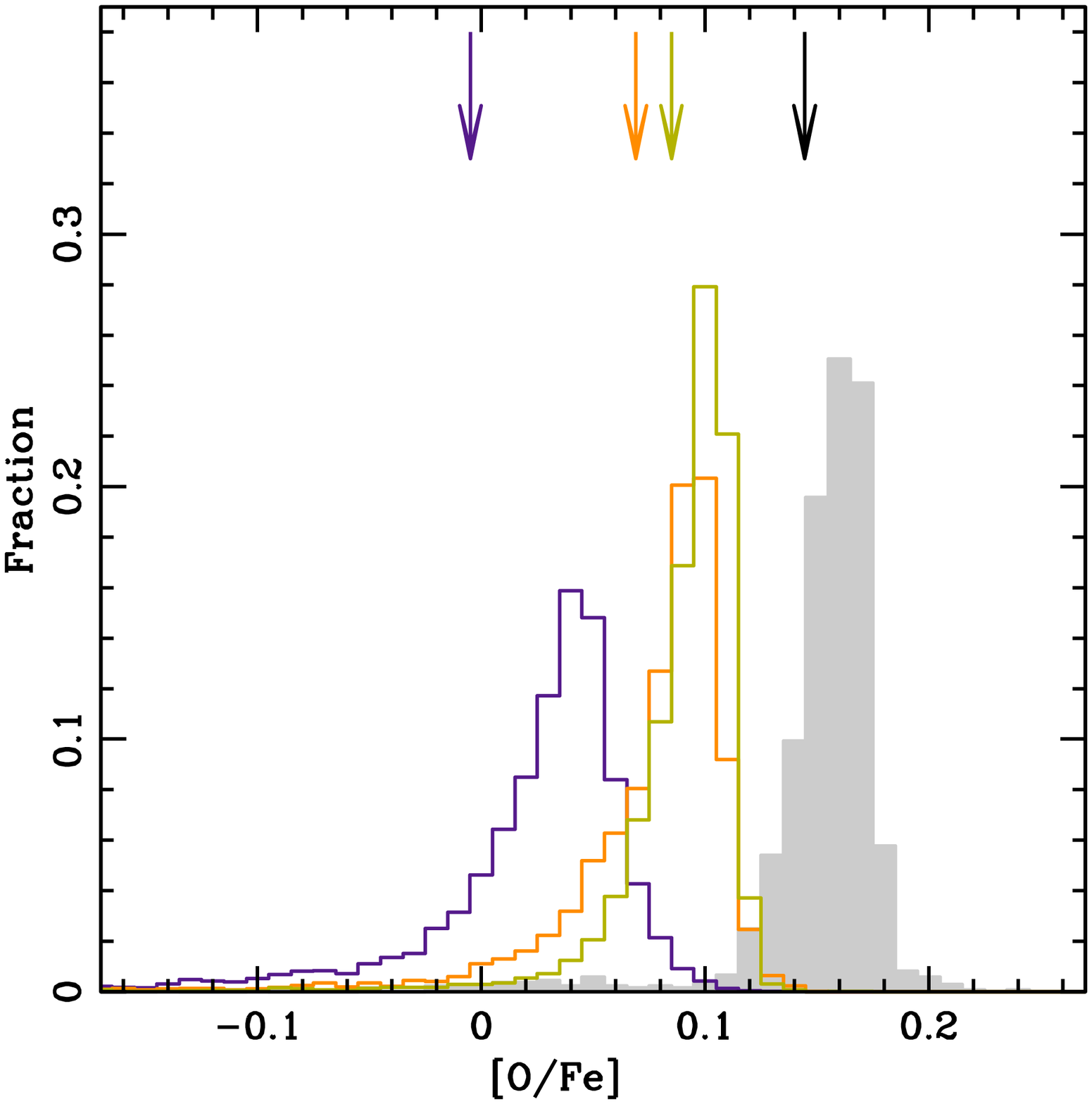,width=0.33\textwidth} 
 \caption{As in Fig.~\ref{sfr_MW}, but this time showing the average [Fe/H]
   (top panel) and [O/Fe] (bottom panel) of stars in the disk of Milky
   Way-like galaxies.}        
{\label{metal_MW}} 
\end{figure}

As discussed in Section~\ref{Chemistry}, our chemical evolution model  
cwas tuned to reproduce the distribution of Fe in the Milky Way disk,
using simulations from the Aquarius project. When running the same
model on a cosmological volume, we found that a few parameters needed  
to be slightly modified in order to match the normalisation of the
mass-metallicity relation observed in the local Universe.
Specifically,  we had to reduce both the feedback efficiency (from
$\epsilon_{\mathrm{re-heat}} = 0.05$ to $0.02$), and the
re-incorporation factor (from $\gamma=0.5$ to $0.1$).  

To verify how much these parameter changes affect the results
discussed in our  previous work, we have used the Millennium
simulation to select Milky Way-like galaxies using criteria as close
as possible to those used to select the Aquarius
haloes. Specifically, we have considered only central galaxies of
haloes with mass between $\sim 6.5\times10^{11}\, {\rm M}_{\odot}$ and
$\sim 1\times10^{12}\,{\rm M}_{\odot}$ with no halo within a sphere of
1 Mpc with mass larger than half its mass. 

Fig.~\ref{sfr_MW} shows the distribution of stellar masses, cold gas 
masses and star formation rates obtained for galaxies selected using
these criteria in different models. The grey shaded histogram, in
particular, refers to the model used in \citet{DeLucia14}, while the
coloured histograms correspond to the fiducial (lila), FIRE (orange),
and zDEP (yellow) models used in this paper. The vertical dotted lines
mark the observational estimates as given in  \citet{DeLucia14}, and
the arrows indicate the median of the distributions.  Considering the
FIRE and zDEP models, that are our favourite schemes on the basis of
the results presented in this paper, we have on average slightly lower
stellar masses, larger cold gas masses, and larger star formation
rates with respect to results from the model used in
\citet{DeLucia14}.  We note, however,  that there is still a sizeable
population (comparable in number to that found on the basis of the
original model used in \citealt{DeLucia14}), whose physical properties
are comparable to those estimated for our Galaxy.  

Fig.~\ref{metal_MW} shows the distributions of [Fe/H] and [O/Fe]
obtained for  the Milky Way-like galaxies selected as discussed above,
for the same models. In this case the changes are more dramatic (not
surprisingly as we needed to introduce these modifications to change
the overall normalisation of the mass-metallicity relation in the
first place): both the FIRE and zDEP models (as well as our fiducial
model) predict average values for the [Fe/H] of Milky Way disks that
are larger than those obtained in \citet{DeLucia14}. The opposite is
true for [O/Fe], although here the distribution measured for the stars
in our Galaxy disk is very large and the median might be not very
informative. We plan, in future work, to apply our new favourite
scheme to the Aquarius simulations and reconsider the chemical
properties of both the Milky Way and its satellites. 

%%%%%%%%%%%%%%%%%%%%%%%%%%%%%% Yates et al. criteria %%%%%%%%%%%%%%%%%%%%%%%
\begin{figure}
 \centering \vspace{1.0cm}
 \epsfig{file=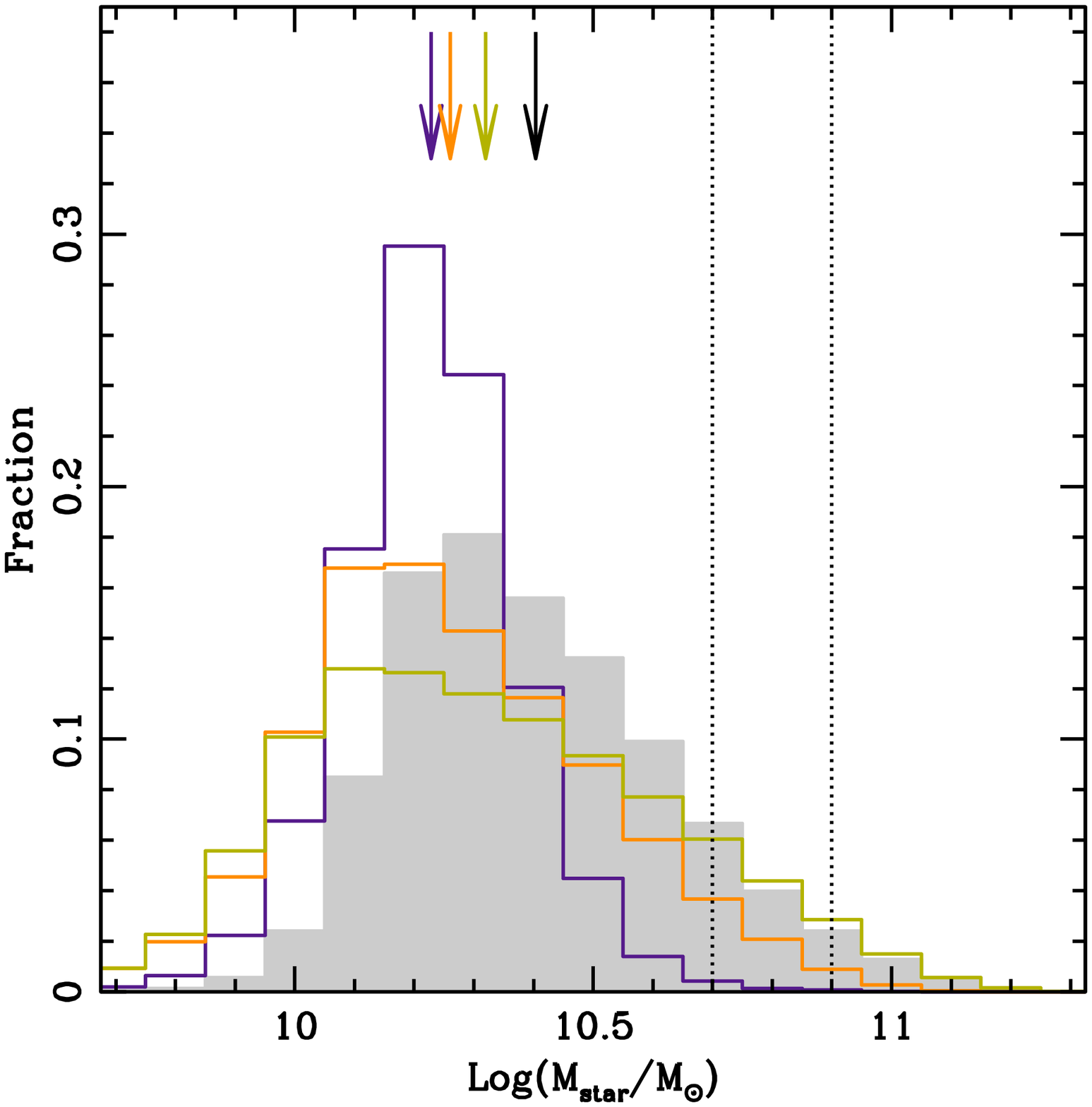,width=0.33\textwidth}\vspace{1.5cm}
 \epsfig{file=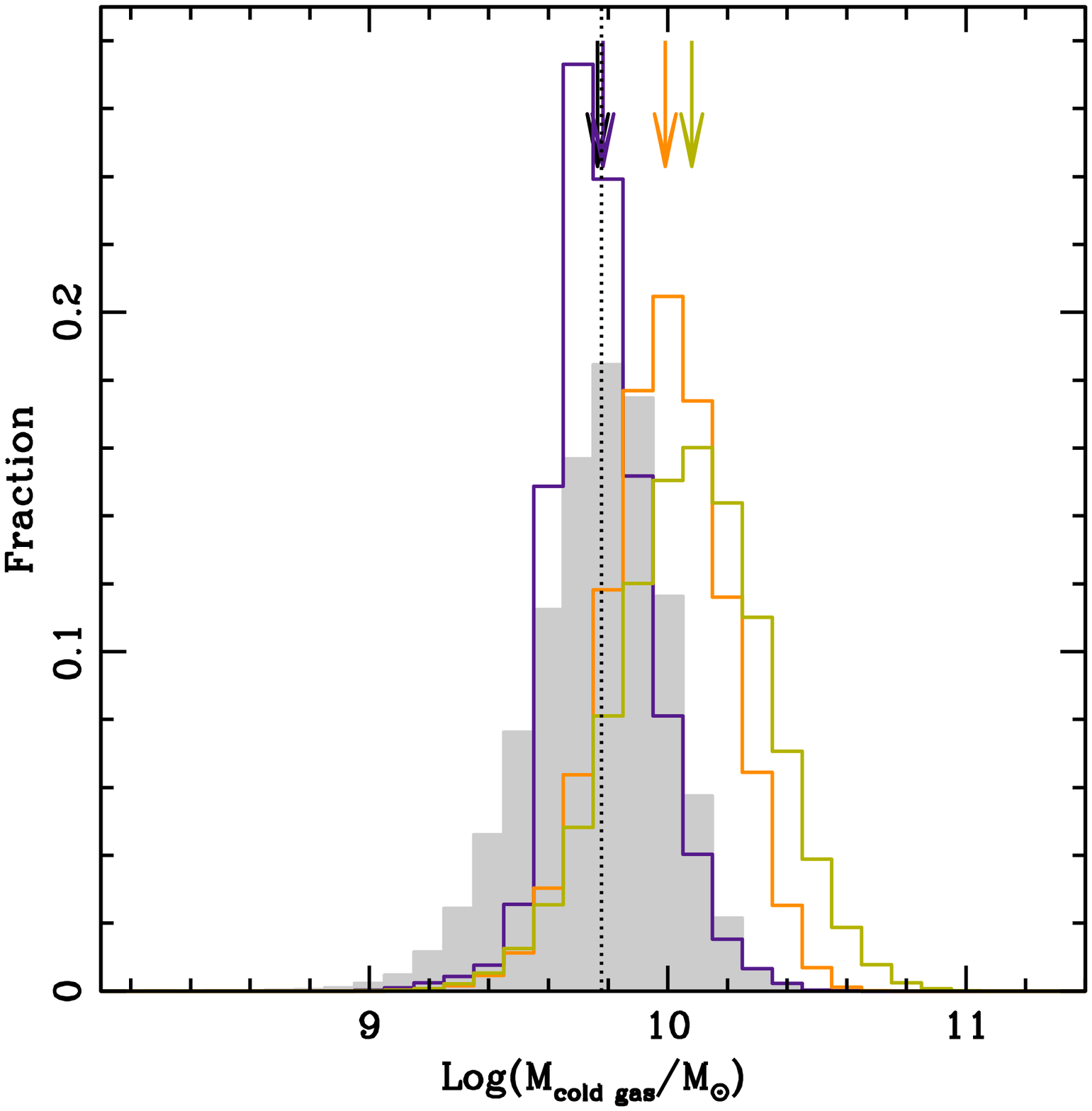,width=0.33\textwidth}\vspace{1.5cm}
 \epsfig{file=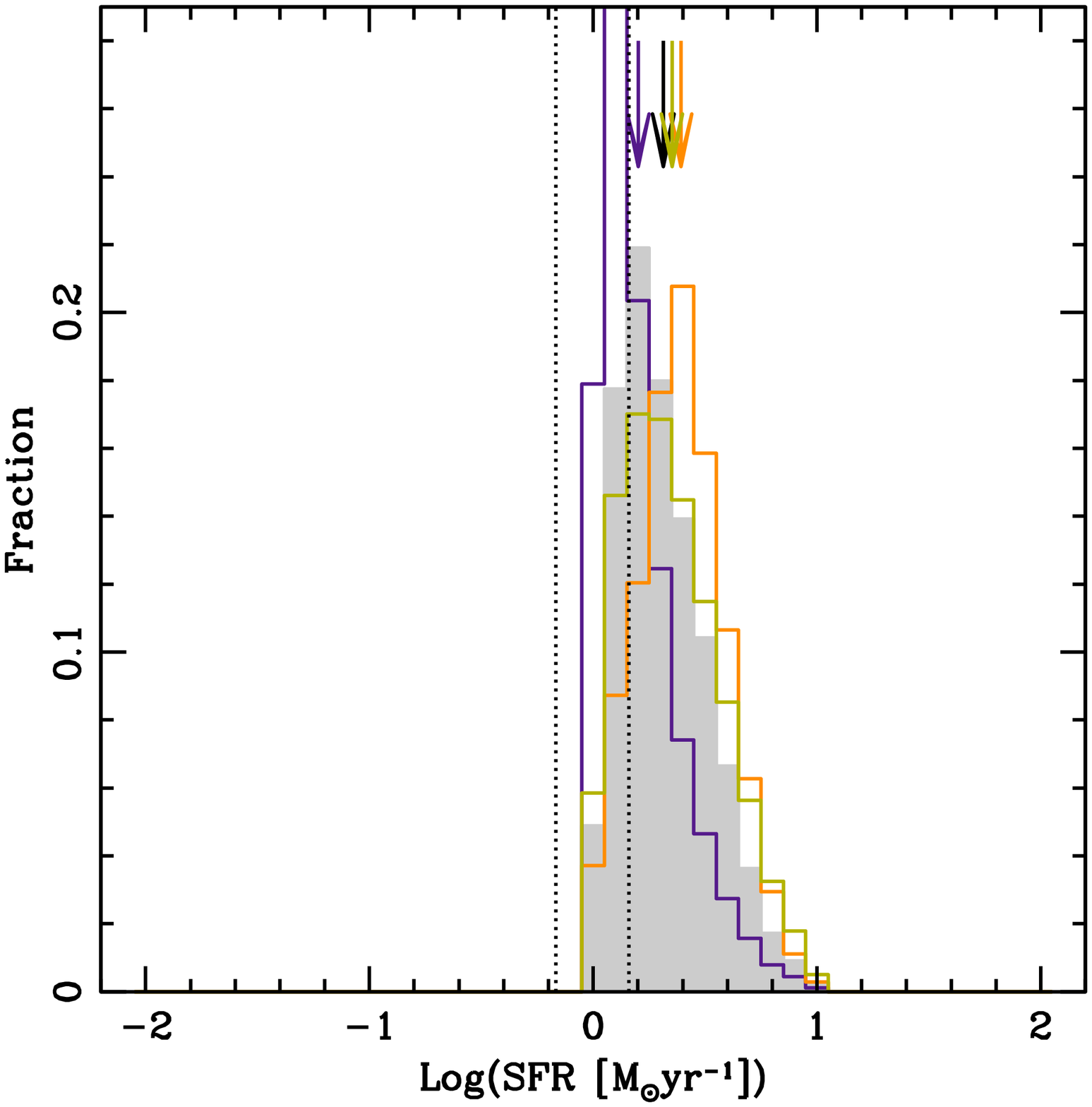,width=0.33\textwidth} 
 \caption{As in Fig.~\ref{sfr_MW}, but using the Yates-like criteria described
   in the text.}   
\label{MWpropY}     
\end{figure}

\begin{figure}
 \centering \vspace{1.0cm}
 \epsfig{file=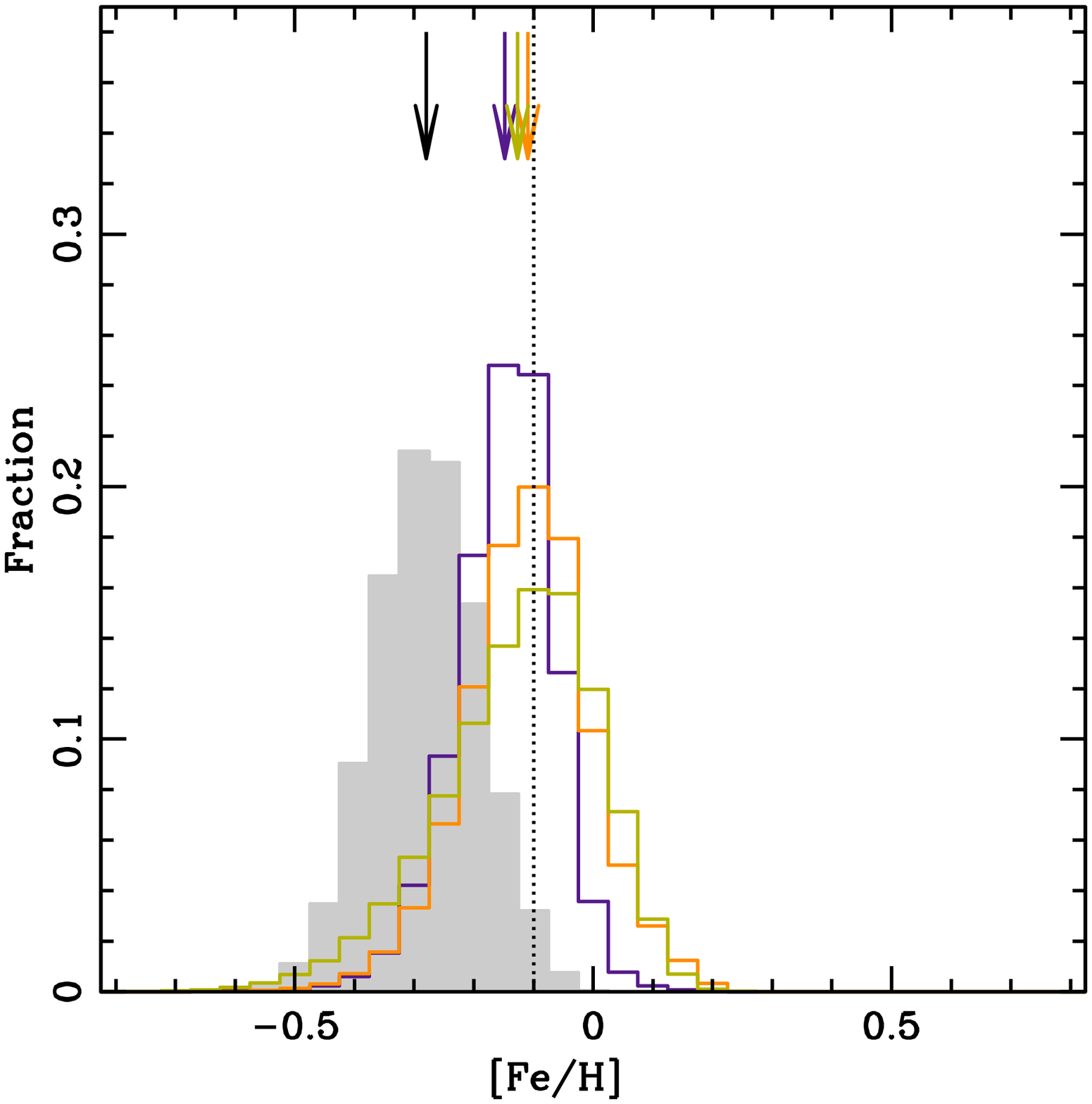,width=0.33\textwidth}\vspace{1.5cm}
 \epsfig{file=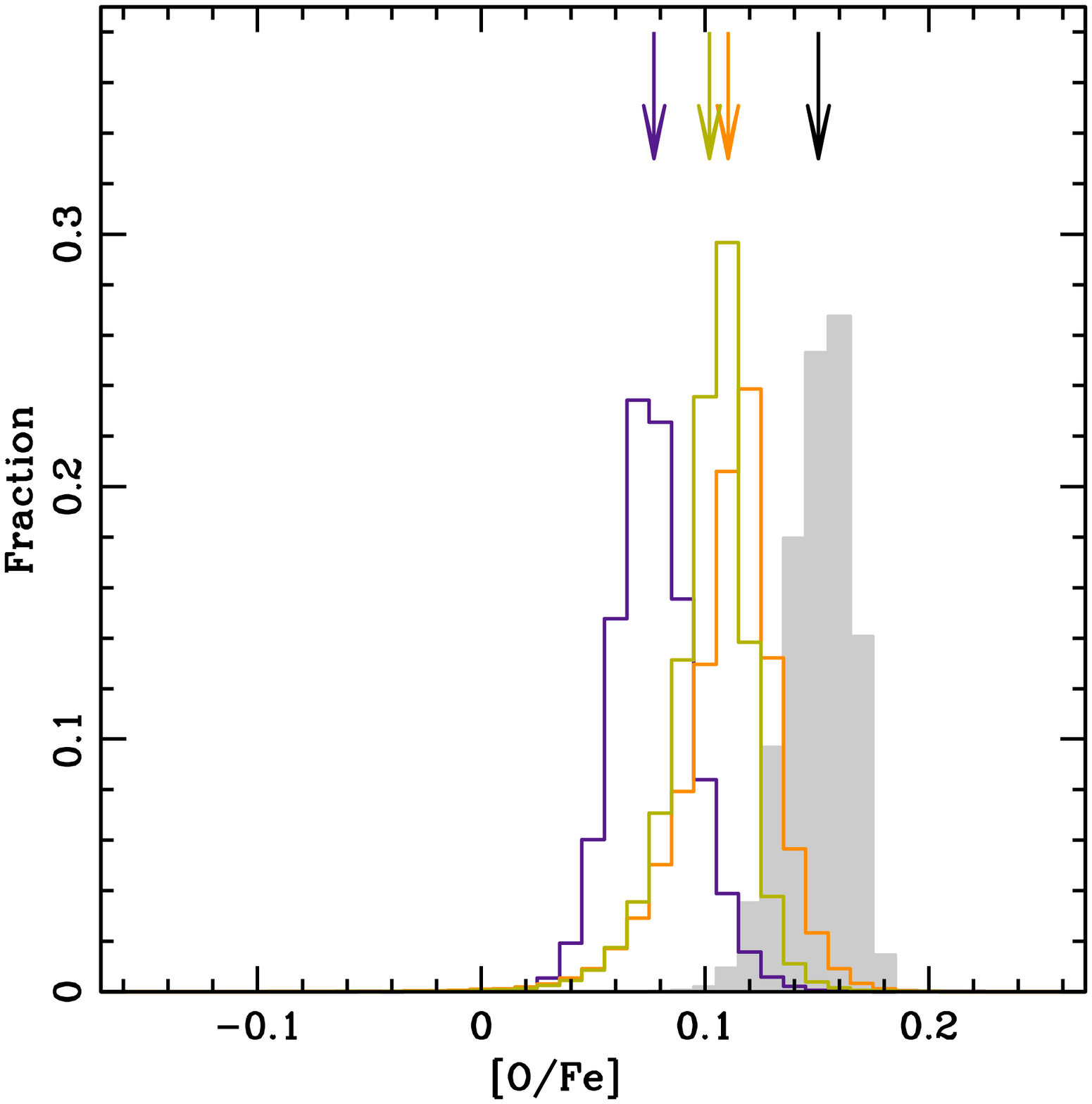,width=0.33\textwidth} 
 \caption{As in Fig.~\ref{metal_MW}, but using the Yates-like criteria
   described in the text.}
\label{MWmetY}
\end{figure}

In a recent paper, \citet{Yates13} have included an independently
developed chemical enrichment model into the semi-analytic model by
\citet{Guo11}. They show that their implementation reproduces,
simultaneously, the chemical properties of Milky Way-like galaxies and
the observed mass-metallicity relation in the local Universe. We note
that they do not discuss how their updates affect the global
properties of the galaxy population in the local Universe and, in
particular, they do not show the galaxy stellar mass function. Their
Milky Way-like galaxies are selected from the Millennium, but using
criteria slightly different from those we have used above. In
Figs.~\ref{MWpropY} and \ref{MWmetY}, we show the same distributions
considered above, but this time using the same criteria adopted in
\citet{Yates13}. Specifically, we select only central galaxies in
haloes with mass between $\sim 3\times10^{11}\,{\rm M}_{\odot}$ and
$\sim 3\times10^{12}\,{\rm M}_{\odot}$, SFR measured below $z=0.25$
between $1\,{\rm  M}_{\odot}{\rm yr}^{-1}$ and $10\,{\rm
  M}_{\odot}{\rm yr}^{-1}$, and stellar bulge-to-total ratio smaller
than 0.5. When considering the FIRE and zDEP models, these criteria
bring the cold gas mass and SFR (by construction) of our Milky
Way-like galaxies much closer to the observational estimates. This
applies also for the average [Fe/H] of stars in the Milky Way
disks. Therefore, our favourite models are able to reproduce,
simultaneously, the metal content of Milky-Way like galaxies and the
overall normalisation of the mass-metallicity relation of local
galaxies (as for the model discussed in \citealt{Yates13}), as well as
the GSMF in the local Universe and its evolution up to $z\sim3$.
%%%%%%%%%%%%%%%%%%%%%%%%%%%%%%%%%%%%%%%%%%%%%%%%%%%%%%%%%%%%%%%%%%%%%%%%%%%%%

\section{Resolution study}\label{resolution}

% -- first of all, I am not sure what the vertical line is meant to show in
% the figure showing the baryon conversion efficiency. You should clarify 
% it (I think the numbers you had quoted in the text were incorrect because
% you had not included h=0.73?). This actually applies also to the figure
% in the main text.

% -- As for the convergence at the massive end, I am a bit puzzled as if you
% argue it is a problem of low number statistics, then it should be even more
% evident at higher redshift.... Can this be due to the fact that now the 
% satellites are more gas rich and so you have more gas rich mergers onto 
% central galaxies?

% -- For the baryon conversion efficiencies, we should at least try to explain
% why results are different (lines are not exactly on top of each other, at 
% all halo masses, i.e. there is a very weak convergence in this case). At 
% the very massive end, this is the same problem of over-producing the massive
% galaxies?

In this section, we discuss to what extent our model predictions vary
when increasing the resolution of the underlying dark matter
simulation. To this aim, we take advantage of the MillenniumII
simulation (\citealp{Boylan-Kolchin09}), that adopts the same
cosmology as the Millennium simulation, but corresponds to a smaller
box (100 Mpc h$^{-1}$ against 500 Mpc h$^{-1}$), five times better
spatial resolution (the Plummer equivalent softening of the Millennium 
II is 1.0 kpc h$^{-1}$) and 125 times better mass resolution (the
particle mass of the Millennium II simulation is $6.9 \times 10^6
M_\odot h^{-1}$).  

\begin{figure*}
 \centering
 \epsfig{file=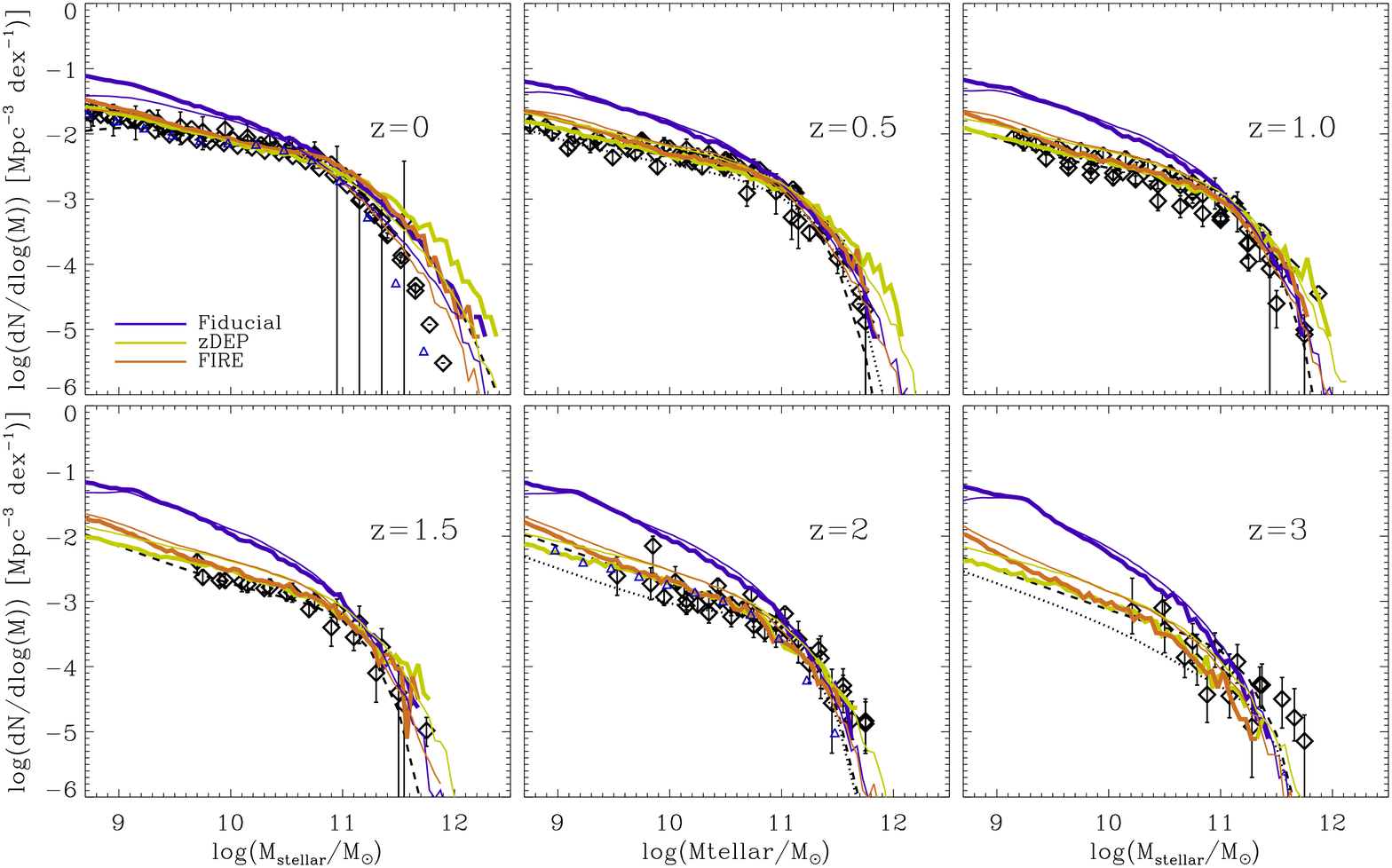,width=0.8\textwidth}
 \caption{Evolution of the GSMF for the fiducial, zDEP and FIRE models based on
   the Millennium (thin solid lines), and on the Millennium-II trees
   (thick solid lines), and compared to observational measurements
 (black symbols and black  lines). } {\label{SMF_II}}
\end{figure*}

Fig. \ref{SMF_II} shows the evolution of the galaxy stellar mass
function for the Fiducial, zDEP, and FIRE models based on the
Millennium (thin solid lines) and the MillenniumII simulation (thick solid
lines), and compares model predictions to observational measurements 
(as discussed in Fig. \ref{SMF_evol}). For galaxies with masses
  between $10^9-10^{11} M_\odot$ in the FIRE and zDEP models, we
  find small differences ($<0.2$~dex) between the different resolution
  runs at high redshift $z>0$. At redshift $z=0$, some discrepancies
are visible at the very massive end for all models, and at the
  low mass end ($<10^{10} M_\odot$) for the fiducial model by maximum
  0.2~dex. The former can be at least in part due to the small number
statistics for very massive galaxies in the smaller box of the
Millennium II.  In addition, infalling (satellite) galaxies tend to be
more gas rich in the higher resolution simulation, leading to more gas
rich mergers onto central galaxies and thus, to more star formation.

\begin{figure*}
 \centering
 \epsfig{file=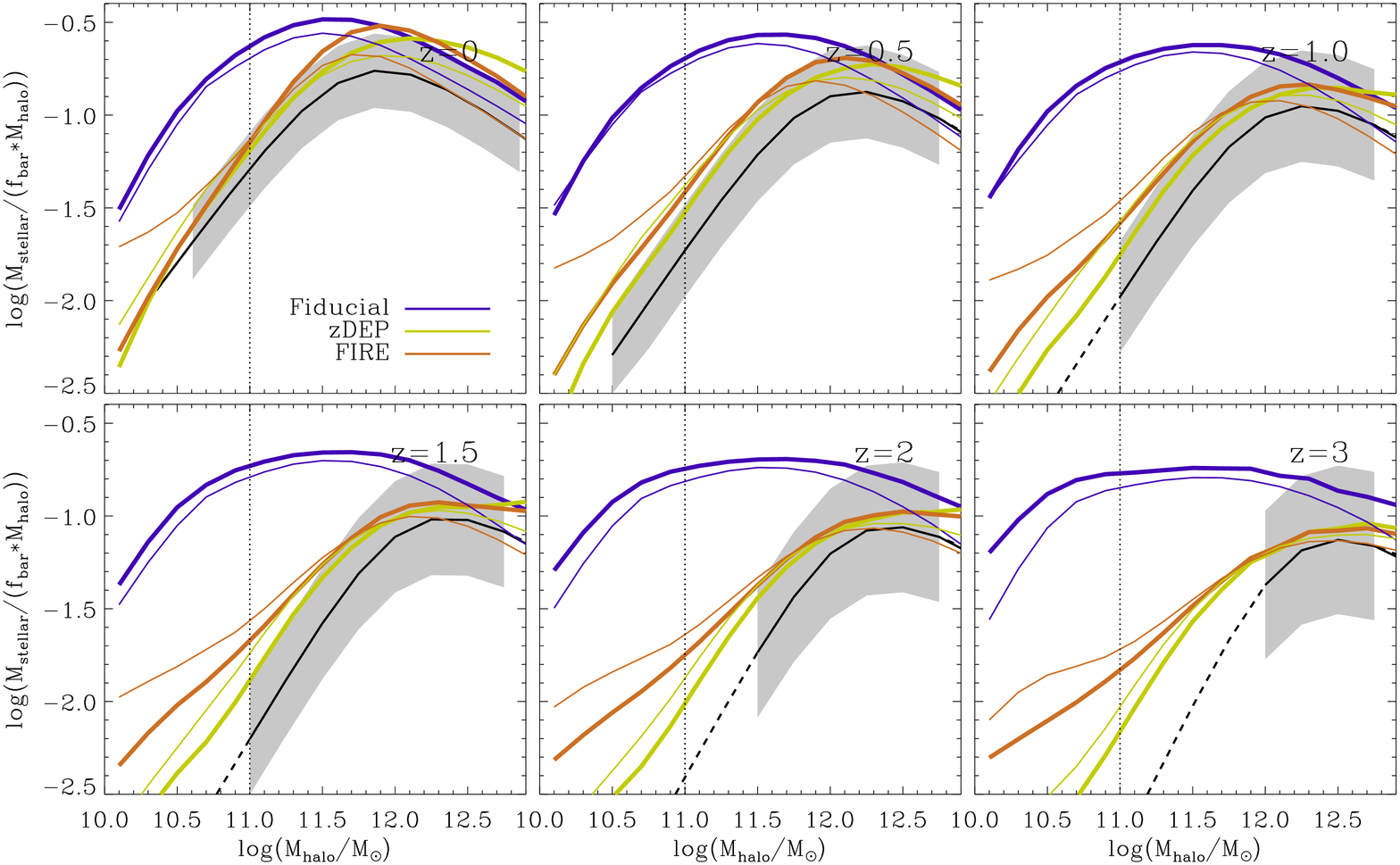,width=0.8\textwidth}
 \caption{ Evolution of the average baryon conversion
   efficiencies for the fiducial,  zDEP and FIRE models based on the
   Millennium (thin solid lines) and on the 
   Millennium-II trees (thick solid lines), compared to predictions from subhalo
   abundance matching methods (black lines with grey shaded
   areas). The black dotted line indicates the mass limit, where model
 predictions start to get more strongly affected by resolution.}
         {\label{BarConv_II}}
\end{figure*}

Fig. \ref{BarConv_II} illustrates the evolution of the baryon
conversion efficiencies as a function of halo mass for the fiducial,
the zDEP, and the FIRE models, based on merger trees from
the Millennium (thin solid lines) and the MillenniumII (thick solid lines)
simulation. A 1000 particle, and thus resolved, halo in the Millennium
simulation corresponds to a mass of $\sim 1 \times 10^{12} M_\odot
h^{-1}$, and $ \sim 9 \times 10^{9} M_\odot h^{-1}$ for the
MillenniumII simulation.

We find that baryon conversion efficiencies only weakly converge
with resolution, even if for a halo mass range of $10^{11}-10^{12}
M_\odot$ the differences are small. Conversion  efficiencies of
massive halos in the higher resolution simulation are larger, which is
a reflection of the larger number densities of massive galaxies in the
Millennium-II run compared to the Millennium simulation as shown in
Fig. \ref{SMF_II}. 

At $z=0$ and $z=5$, the FIRE and zDEP predictions, based on the better
resolved merger trees, are still in fairly good agreement with
abundance matching predictions (grey shaded areas,
\citealp{Moster13}). At higher redshifts, abundance matching
predictions are not available for halo masses below $10^{11}
M_\odot$. When extrapolating their fits, baryon conversion
efficiencies of the high-resolution FIRE and  zDEP models,  which are
slightly reduced compared to the lower-resolution models, are in even
better agreement with the abundance matching trends.

\end{appendix}

\end{document}